\documentclass[12pt]{article}
\usepackage[left=30mm, right=30mm]{geometry}
\renewcommand{\baselinestretch}{2}
\DeclareMathAlphabet{\pazocal}{OMS}{zplm}{m}{n}
\newcommand{\Lb}{\pazocal{X}}

\usepackage[position=top]{subfig}
\usepackage{bm}
\usepackage{dsfont}
\usepackage{amsmath}
\usepackage{url}
\usepackage{array}
\usepackage{nicefrac}
\usepackage{float}
\usepackage{longtable}
\usepackage{afterpage}
\usepackage{graphicx}
\usepackage{epstopdf}
\usepackage[singlelinecheck=off]{caption}

\usepackage[utf8]{inputenc}
\usepackage[english]{babel}
\usepackage[sectionbib,square]{natbib}

\usepackage{booktabs}

\usepackage[flushleft]{threeparttable}
\usepackage{multicol}
\usepackage{multirow}
\usepackage{tablefootnote}
\usepackage{fourier}

\usepackage{scrextend}
\addtokomafont{labelinglabel}{\sffamily}
\usepackage{lscape}
\usepackage[flushleft]{threeparttable}

\usepackage[skins,breakable]{tcolorbox}
\usepackage{enumitem}

\begin{document}

\begin{center}
\Large{An agglomerative hierarchical clustering method by optimizing the average silhouette width}
\end{center}
\begin{center}
\noindent\textbf{ Fatima Batool}\\
 ucakfba@ucl.ac.uk \\
Department of Statistical Science, \\
University College London, WC1E 6BT, United Kingdom
 \end{center}   
\begin{center}
\textbf{Abstract} 
\end{center}
\noindent An agglomerative hierarchical clustering (AHC) framework and algorithm named HOSil based on a new linkage metric optimized by the average silhouette width (ASW) index ( \cite{rousseeuw1987silhouettes}) is proposed.  A conscientious investigation of various clustering methods and estimation indices is conducted across a diverse verities of data structures for three aims: a) clustering quality, b) clustering recovery, and c) estimation of number of clusters. HOSil has shown better clustering quality for a range of artificial and real world data structures as compared to  $k$-means, PAM, single, complete, average, Ward, McQuitty, spectral, model-based, and several estimation methods.  It can identify  clusters of various  shapes including spherical, elongated, relatively small sized clusters, clusters coming from different distributions including uniform, t, gamma and others. HOSil has shown good recovery for correct determination of the number of clusters. For some data structures only HOSil was able to identify the correct number of clusters. 

\section{Introduction}
\noindent Clustering is a widely used multivariate data analysis tool that aims at uncovering hidden underlying grouping structure in the data. It has become  an essential data analysis procedure in various disciplines for different purposes, for instance, to organize the big, messy datasets for further analysis,  to find important features, to explore the relationship between features, or to reduce dimensions. Some application areas include but are not limited to medical science, neuroscience, climatology, data mining, and computer vision. 

\par In this article we have solved two major problems in cluster analysis simultaneously. The two problems are the estimation of number of clusters and finding the clustering solution. A clustering method can be defined by optimizing the objective function based on the ASW clustering quality index proposed by \cite{rousseeuw1987silhouettes}.  The motivation is that if the number of clusters estimated by an index are acceptable then the final clustering solution based on the criterion used by the index should also be acceptable. The advantage of this is that it will make the task of clustering somewhat simpler and straightforward and the users don't have to deal with the two tasks separately. In particular, in this article the AHC framework has been considered based on the  optimizing of the ASW. The new algorithm proposed based on this method is named as HOSil (Hierarchical Optimum Silhouette width).  

\par Clustering methods can be broadly classified as the hierarchical methods and the non-hierarchical methods.  The hierarchical methods return a sequence of clustering solutions whereas non-hierarchical, also known as  partitional or flat clustering, returns a single clustering solution.  The aforementioned clustering methods have several advantages over the latter, for instance,  for some applications a sequence of clustering solutions might be  more informative as compared to just a single solution. \par Secondly, unlike most non-hierarchical algorithms there is no need to specify the number of clusters a priori,  however, if the desired number of clusters is already known the partitioning can be stopped when the required clusters are obtained. The results of the hierarchical clustering are organized in a dendogram such that the clustering solution at each hierarchy is visible in a single diagram.

\par The standard hierarchical clustering methods are deterministic and can be further classified as agglomerative (bottom-up) and divisive (top-down) methods.  In this work, an AHC framework has been developed  by taking the bottom-up approach. In the beginning of the bottom-up AHC each observation is considered as a ``singleton'' cluster. Based on some  measure of proximity between clusters the process of developing hierarchy by merging the clusters begins. The  measure of proximity commonly known as linkage method can be measured in form of similarities or dissimilarities. We have proposed ASW (\cite{rousseeuw1987silhouettes}) as a linkage measure and used it to merge the clusters.  A distance measure for the inter-cluster pairs of observations used for the calculation of the linkage method, which are the functions of these pairwise distances, are needed to amalgamate clusters. We have used Euclidean distance between pairs of observations. After each merge the total number of clusters decreases by one.

\par After the construction of the dendogram the next task is to decide where to cut the dendogram or in other words decide upon the number of clusters. The advantage of using ASW as a linkage criterion is that it automatically gives the best number of clusters.  The tasks of evaluation of clustering quality and estimation of number of clusters are closely related. Since the ASW is a clustering quality measure the idea here is to choose the number of clusters that give the best clustering quality. The  dendogram is cut at the level where  the maximum ASW value is obtained. 


\paragraph{Organization } In \S \ref{relatedwork} brief literature has been presented to ASW.  In \S \ref{sectionasw} the notations for this paper have been set up and the ASW index has been formalized. In \S \ref{OASWclustering} the optimum ASW based AHC method and algorithm have been proposed. Exploration of the proposed method has then been focused on\textemdash the kind of structures it can  can handle and its comparisons with other methods. In \S \ref{simulationsetup} experiments have been set up. The performance of the HOSil algorithm has been evaluated in two respects a) for the estimation of number of clusters, b) for the clustering solution itself. The references to all the clustering methods, indices for the estimation of number of clusters, and their software implementation used in this work have been presented in \S \ref{simulationsetup}. The simulation results have been presented in \S \ref{applications}.  In \S \ref{runtimeanalysis} the numerical complexity analysis of the HOSil has been considered. Real-life applications have been presented in \S \ref{simres}. Finally, the study has been closed with conclusions and future directions for this work in \S \ref{con}. This article has an appendix and a supplementary file.


\section{Related work} \label{relatedwork}

\noindent \cite{rousseeuw1987silhouettes} has proposed ASW with the partitioning around medoid algorithm (PAM) to estimate number of clusters. The ASW is a well-reputed and trusted clustering quality measure. The index has been well received by the research community and is widely used for the estimation of the number of clusters.  The ASW has been extensively used to estimate the optimal number of clusters  (with a combination of various clustering methods), to compare the performance of clustering methods and for the quality assessment of clustering obtained from many clustering methods. Some empirical  studies have also been designed to evaluate performance of the ASW in comparison with other famous indecies. Some examples include  \cite{recupero2007new}, \cite{kennedy2003large},  \cite{hruschka2003genetic} and \cite{lovmar2005silhouette}. For clustering quality measures, and clustering method comparisons see  \cite{liu2003algorithms},  \cite{reynolds2006clustering}, \cite{ignaccolo2008analysis} and \cite{arbelaitz2013extensive}.  \cite{campello2006fuzzy} have extended ASW to a fuzzy clustering regime. Some interesting variations and modifications have also been proposed, for instance, density based ASW by \cite{menardi2011density} and the slope statistics by \cite{fujita2014non}. 

\par  \cite{van2003new} have proposed PAMSIL algorithm which is a PAM-like (\cite{rousseeuw1987silhouettes}) algorithm to maximize the ASW using medoids to find flat clustering. The PAM algorithm has two phases, namely, the build phase, and the swap phase. \cite{van2003new} first ran the build phase of PAM algorithm to get  a set of $k$-medoids and then considered all possible swaps to further improve the values of objective function obtained in the build phase with an exception that their objective function did not try to optimize the distance of points from the nearest medoid but the ASW of the clustering. 

\section{Notational setup } \label{sectionasw}
  The silhouette width (SW) for an object in data represents how well the object fits in its present cluster. Let $\Lb = \{x_1, \dots, x_n\}$  be the data of size $n$ and $d$ be a distance function over $\Lb$ and $\mathcal{C}_k = \{C_1, \dots, C_k \}$ a clustering identified by some clustering function $f_k$ on $\Lb$. Let $i$ represents the index for observations $x_i \in \Lb$.    Let the clustering labels set be denoted by $\{l(1), \dots, l(n)\} \in \mathds{N}_k$ determined by $l(i) = r$, $r \in \mathds{N}_k$, $i \in \mathds{N}_n$ and cluster sizes are determined by $n_r = \sum_{i=1}^{n} 1 (l(i) = r)$, $r \in \mathds{N}_k$. For each objects  $i \in \mathds{N}_n$ calculate
\begin{equation} 
a(i) =  \frac{1}{n_{{l(i)}}-1} \sum_{\substack{  l(i) = l(h)\\
                 i \neq h }} d(x_i, x_h),   \quad\text{and}\quad b(i) =  \min_{r \neq l(i)} \frac{1}{n_r} \sum_{l(h) = r} d(x_i, x_h).
\end{equation}
 \noindent For a given clustering $\mathcal{C}_k$, the silhouette width for a data object having index $i$, $ i \in \mathds{N}_n$, is
\begin{equation} \label{Sbarck}
S_i(\mathcal{C}_k, d) = \frac{b(i)  - a(i)}{ \textrm{max} \{ a(i), b(i)\}},
\end{equation}
such that $-1 \leq S_i(\mathcal{C}_k, d) \leq 1$. 
\par The SW averaged over all the members of a cluster can be used as a measure of a cluster's quality. The ASW averages SW over all members of a dataset $\Lb$. It is a global quality measure for clustering. Formally, for the clustering $\mathcal{C}_k$ it can be written as follows: 
\begin{equation} \label{sbar}
\bar{S}(\mathcal{C}_k, d) = \frac{1}{n}\sum_{i = 1}^n S_i(\mathcal{C}_k, d).
\end{equation}
The best $k$ can be selected by maximising $\bar S(\mathcal{C}_k, d)$ over $k$.
\par The ASW can be thought of as a combinational index because it is based on two concepts which are separation and compactness that define a unified concept of isolation. It is a ratio of inter-cluster variation and  intra-cluster variation.  It measures how homogeneous the clusters are, and what the separation between them is. Thus the ASW tells us about the coherent structure of clustering.

\par For a good clustering the  dissimilarity ``within'' clusters should  be less than the dissimilarity  ``between'' clusters. Therefore, if $a(i)$ is much smaller than the smallest between clusters dissimilarity $b(i)$ we get evidence (larger $s(i)$, close to 1 is better in this case),  that object $x_i$ is in the appropriate cluster. On the other hand, $s(i)$ close to -1, points towards the wrong cluster assignment for object $x_i$. In this case $a(i) > b(i)$,  meaning that object ``$i$'' is more close to its neighbouring cluster than to its present cluster. A neutral case occurs when  $s(i) \approx 0$, i.e., object $x_i$ is approximately equidistant from both, its present cluster and neighbouring cluster.


\section{AHC based on optimum ASW linkage   } \label{OASWclustering}
\noindent Let $\Lb = \{x_1, \dots, x_n\}$ be the dataset to partitioned, where $x_{i}$ represents the $i^{th}$ observation, and each $x_{i}$ represents a  $p$-dimensional variable. We will here only consider crisp clustering, thus every object will belong to one cluster only and there will be no overlapping between clusters in the hierarchy. There will be $n$ total hierarchy levels. Let $k_1, \cdots, k_n$ be the number of clusters in a clustering at each hierarchy level. Let the full hierarchy of $\Lb$  is given by $\mathcal{P} = \{ \mathcal{C}^1_{n}, \dots, \mathcal{C}^n_{1}\}$. The superscript in  $\mathcal{C}^l_{k_l}\in \mathcal{P}$, where $l = 1, \cdots, n$ represents the hierarchy level, and $k_l = n, (n-1), \cdots, 2, 1$ represents the number of clusters at each hierarchy level.  In AHC we start with $n$ clusters in the beginning. Thus if $l$ represents a particular hierarchy level, then at $l=1$ we have $k_l=n$ clusters, i.e., each observation forms a separate cluster.  The number of clusters subsequently reduce as hierarchy  level proceeds. For simplicity assume that only one pair of clusters merges at each hierarchy level.

\par Let $\mathcal{C}^l_{k_l}= \{ C^l_1, \cdots, C^l_{k_l}\}$, where $C^l_{r} \in \mathcal{C}^l_{k_l}$, $r = 1, \cdots, k_l$ represents an  $r$-th cluster in a clustering at hierarchy level $l$. The members of a cluster at hierarchy level $l = 1$ can be further written as $C_1^1 = \{x_1\}$, $C_2^1 = \{x_2\}$, $\dots$, $C_{n}^1 = \{x_n\}$, thus $\mathcal{C}_1 =\{ \{x_1\}, \{x_2\}, \dots, \{x_n\} \}$ and at the $(n^{th})$ final  hierarchy level, $\mathcal{C}_{1}^n = \{x_1, x_2, \dots, x_n\}$, such that $\mathcal{C}_1^n = \Lb$. Let $\gamma^l(x_1, r), \dots, \gamma^l(x_n, r)$, where  $r  = 1, \cdots, k_l$ represent the clustering label vector at hierarchy level $l$. At a hierarchy level $l$, $r$ indicates to which cluster observation $x_i$ has been assigned.

\par The AHC clustering algorithms takes an input the pairwise dissimilarities between observations. The pairwise dissimilarities can be calculated using a distance function.  A function $d: \Lb \times \Lb \rightarrow  \mathds{R}^+$, is called a distance on $\Lb$ if, it satisfies the three properties stated as: (a) $\forall$ $x_i, x_h \in \Lb$, where $i, h \in \mathds{N}_n$, $d(x_i, x_h) \geq 0$, (b) $\forall$ $x_i\in \Lb$,  $d(x_i, x_i) = 0$ , distance of an object to itself is zero, and (c)  for $x_i, x_h \in \Lb$, $d(x_i, x_h) = d(x_h, x_i)$.
 
 \par An AHC algorithm has been is defined based on a linkage criterion that optimizes ASW. According to this criterion, two candidate clusters are merged together if this combination gives the maximum ASW  as compared to all other combinations. At each hierarchy level all possible cluster merges have been tried out and those clusters are finally merged that give maximum ASW. HOSil can also be used to find the best number of clusters ($k$) for the data. According to this criterion a best $k$ will be the one that gives maximum value of ASW  among all hierarchy level.

 \par  The algorithm  can't start from $l=1$. This is because for the calculation of $a(i): i \in C$ there should be at least one cluster in the clustering, with at least two observations and for calculation of $b(i)$ there should be at least two clusters in a clustering solution. Therefore, for $(l=1$, $k_1=n)$ and $(l=n$, $k_l=1)$ calculation of ASW is not possible. So we can start calculating ASW from at least $k_2=n-1$ and must stop at $r_{n-1}=2$. For $l=1$ the two closest observations are joined to form a cluster.

\begin{tcolorbox}[float=htpb!, width=\textwidth, colframe=black, arc=0mm, sharp corners=east, colback=white] 
\renewcommand{\baselinestretch}{1}
\textbf{HOSil algorithm}\\
\noindent \rule[2ex]{3cm}{0.80pt} \\
{\fontsize{10}{10} \selectfont
 \textbf{Input}
  \begin{enumerate}
\item[] Input $n(n-1)/2$ pairwise distances between data points in $\Lb$, , i.e., calculate $d(x_i, x_j)$ for all $x_i$, $x_j \in  \Lb$. 
\end{enumerate}
 \textbf{Initialize}
 \begin{enumerate}
\item Set $l = 1$. Start with $n$ clusters i.e., every object forms its own cluster,
\begin{equation*}
\mathcal{C}^l_{k_l} = \Big\{ C^l_1, \dots, C^l_{k_l} \Big\} = \Big\{ \{x_1 \}, \{ x_2 \}, \dots, \{ x_n \} \Big \}, \quad k_l = n.
\end{equation*} 
\item  Update $l=2$. Join the two observations into one cluster that have minimum $d(x_i, x_j)$. Denote the resulting clustering as $\mathcal{C}^l_{k_l} = \Big\{ C^l_1, \cdots, C^\_{k_l} \Big\} $, $k_l = (n-1)$, and  clustering labels for this clustering as $\gamma^l(\Lb, r) = \gamma^l(x_1, r), \dots, \gamma^l (x_n, r)$ where $r = 1, ~ \dots, ~k_l$.
\item Calculate $f^{(l)} = \bar{S}(\mathcal{C}^l_{k_l}, d)$ where $\bar{S}(\cdot)$ as defined in \eqref{sbar}.
\end{enumerate}
\textbf{Repeat}
\begin{enumerate}
\item Combine every cluster $i$  with every other cluster $j$ in the clustering  $\mathcal{C}^l_{k_l}$. For all pairs $(i, j)$ of cluster combinations  denote a set of labels as $\gamma^{l}_{(i, j)}(x_1, r), \dots, \gamma^{l}_{(i, j)} (x_n, r)$, where $r = 1, \cdots, (k_l-1)$ and denote the corresponding clustering as $\mathcal{C}^\ast_{(k_l-1)}$.

\item Calculate $f_{(i, j)} = \bar{S}(\mathcal{C}^\ast_{(k_l-1)}, d) $, where  $\bar{S}(\cdot)$  as defined in \eqref{sbar}. 
\item $(i^\ast, j^\ast)= \text{max} ~~ f_{(i, j)}$, and denote the the corresponding label vector as  $\gamma^l(\Lb, r)$.

\item  Merge the cluster pair ($C^l_{i^\ast}$, $C^l_{j^\ast}$), such that, 
 \begin{equation*}
\bm{\mathcal{C}}^{l+1}_{k_{l+1}} = \mathcal{C}^l_{k_l} \cup \{ C^l_{i^\ast} \cup C^l_{j^\ast} \} \backslash \{ C^l_{i^\ast}, C^l_{j^\ast} \} , \quad k_{l+1} = k_l - 1.
\end{equation*}
Let  $l=l+1$. 

\item Assign $f^{(l)} = f_{(i^\ast, j^\ast)}$.
\end{enumerate}
\textbf{Stop}
 \begin{enumerate}
 \item[] When $l = n - 1$, i.e., $k_l = 2$. 
 \end{enumerate}
\textbf{Return}
 \begin{enumerate}
\item[] $f^{(l)}$ and  $\gamma^l(\Lb, r)$ for all $l = 2, \cdots, (n-1)$.
\end{enumerate}
}
\end{tcolorbox} 

\par
\begin{figure}[!htb]
\center
\subfloat[]{
\includegraphics[width=50mm,height=40mm]{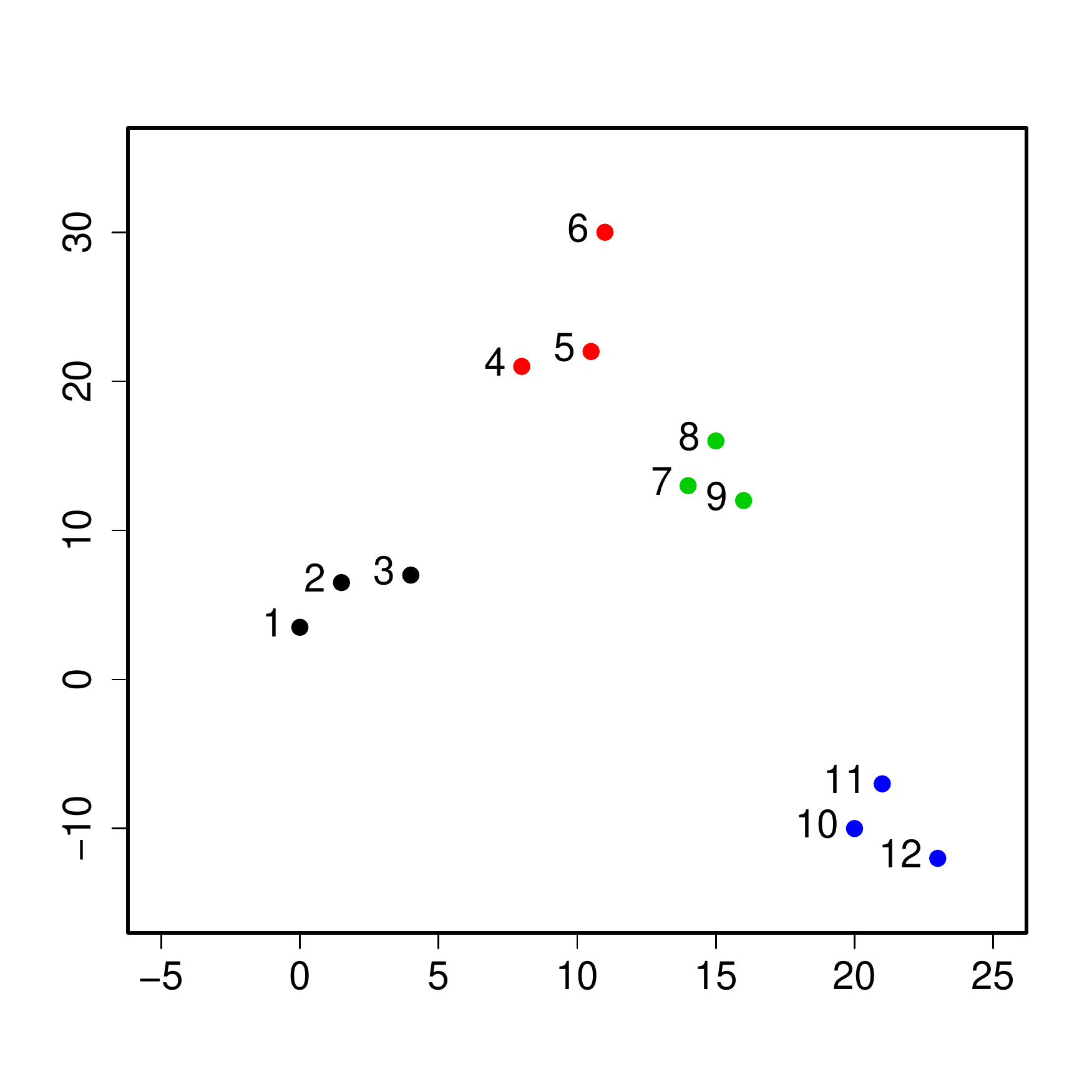}} 
\subfloat[]{
\includegraphics[width=50mm,height=40mm]{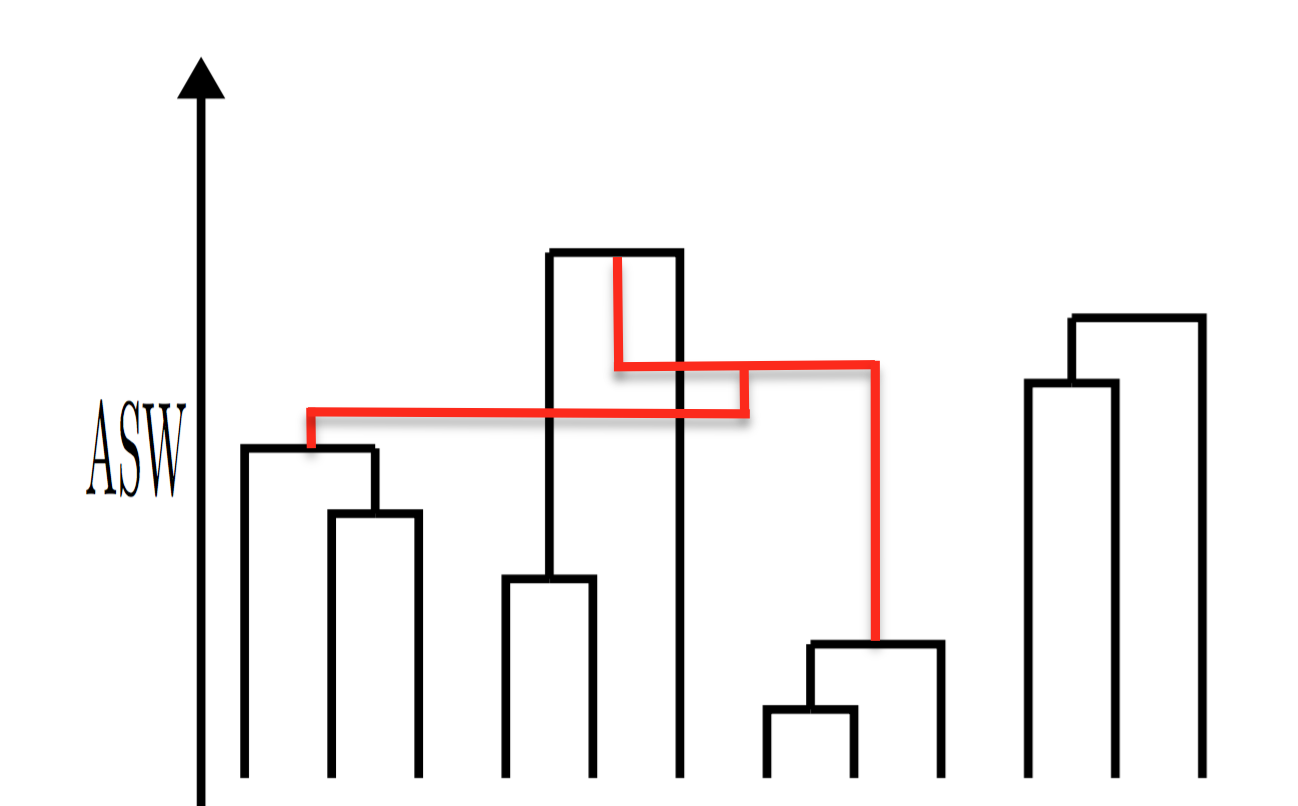}}
\caption{(a) An example with 12 instance in two dimensions to illustrate the HOSil algorithm.. (b) Dendogram representation of the HOSil clustering for the example datasets. The y-axis represents the ASW value obtained from HOSil clustering at each level of the hierarchy.  The node from left to right represents the data points number 1 to 12 plotted in the left panel.}
\label{figureon2}
\end{figure}

 Figure \ref{figureon2}-(a) shows a descriptive example of the algorithm using a dataset of 12 points. The left hand arrow represents the ASW values obtained from HOSil algorithm at each hierarchy level. Based on the maximum ASW principle,  the best number of clusters for the data is 4 which looks intuitive. Using the ASW values obtained on each hierarchy a dendogram has been obtained for this clustering as shown in Figure \ref{figureon2}-(b).  The ASW values obtained at each hierarchy level are plotted on the vertical scale. These values are (k=11, l=1) 0.0613, (k=10, l=2) 0.2014, (k=9, l=3) 0.3177, (k=8, l=4)  0.3811, (k=7, l=5) 0.5049, (k=6, l=6) 0.5496, (k=5, l=7)  0.7117, (k=4, l=8) 0.7218, (k=3, l=10) 0.6408, (k=2, l=11) 0.6080. Note the two red horizontal dendrite connections which represent the decrease in the value of ASW at last two hierarchy level.

\section{Experiments} \label{simulationsetup}
\begin{figure}[!htb]
 \begin{flushleft}
\centering
\subfloat[Model 1]{
\includegraphics[width=35mm,height=35mm]{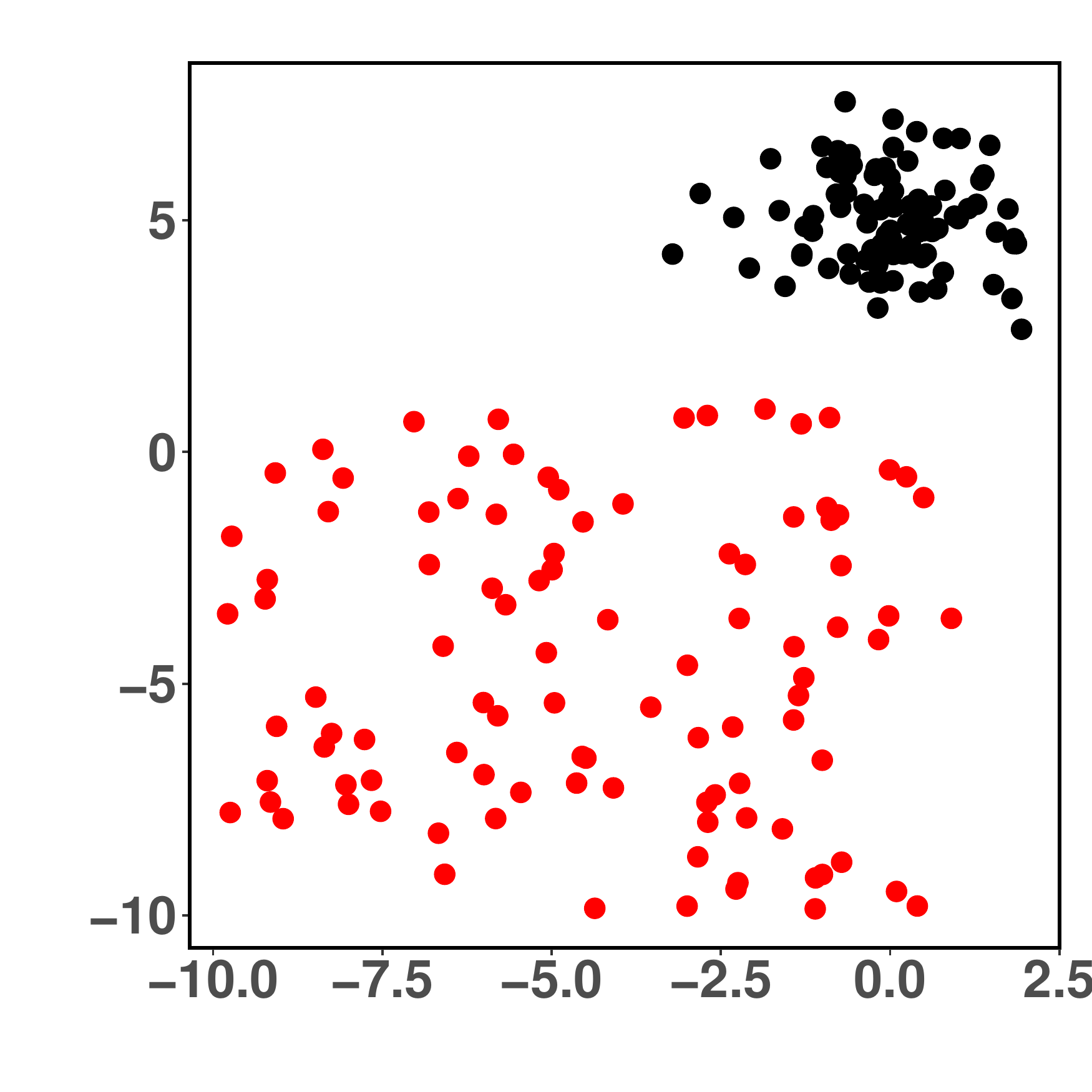} 
}
\subfloat[Model 2]{
  \includegraphics[width=35mm,height=35mm]{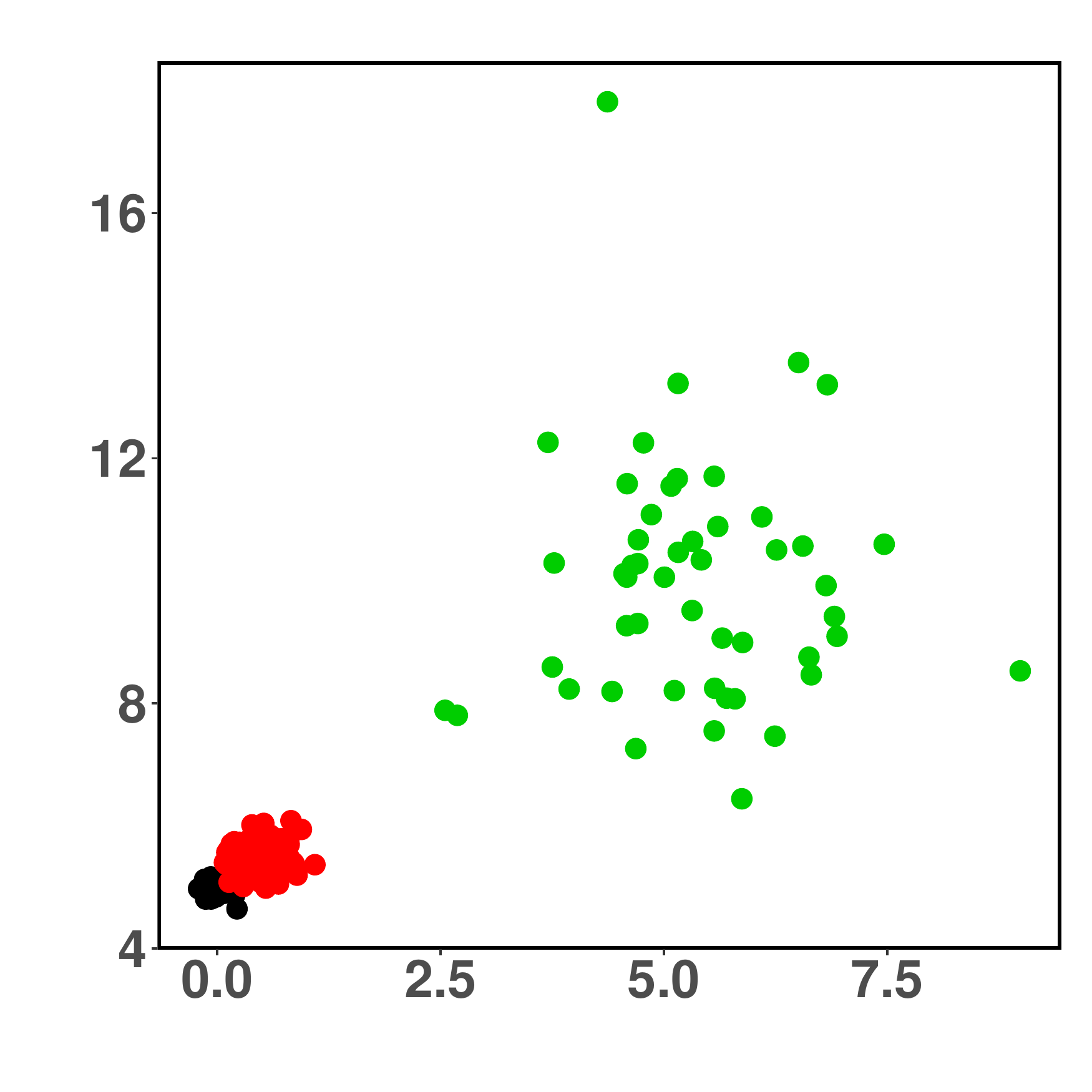}
}
\subfloat[Model 3]{
  \includegraphics[width=35mm,height=35mm]{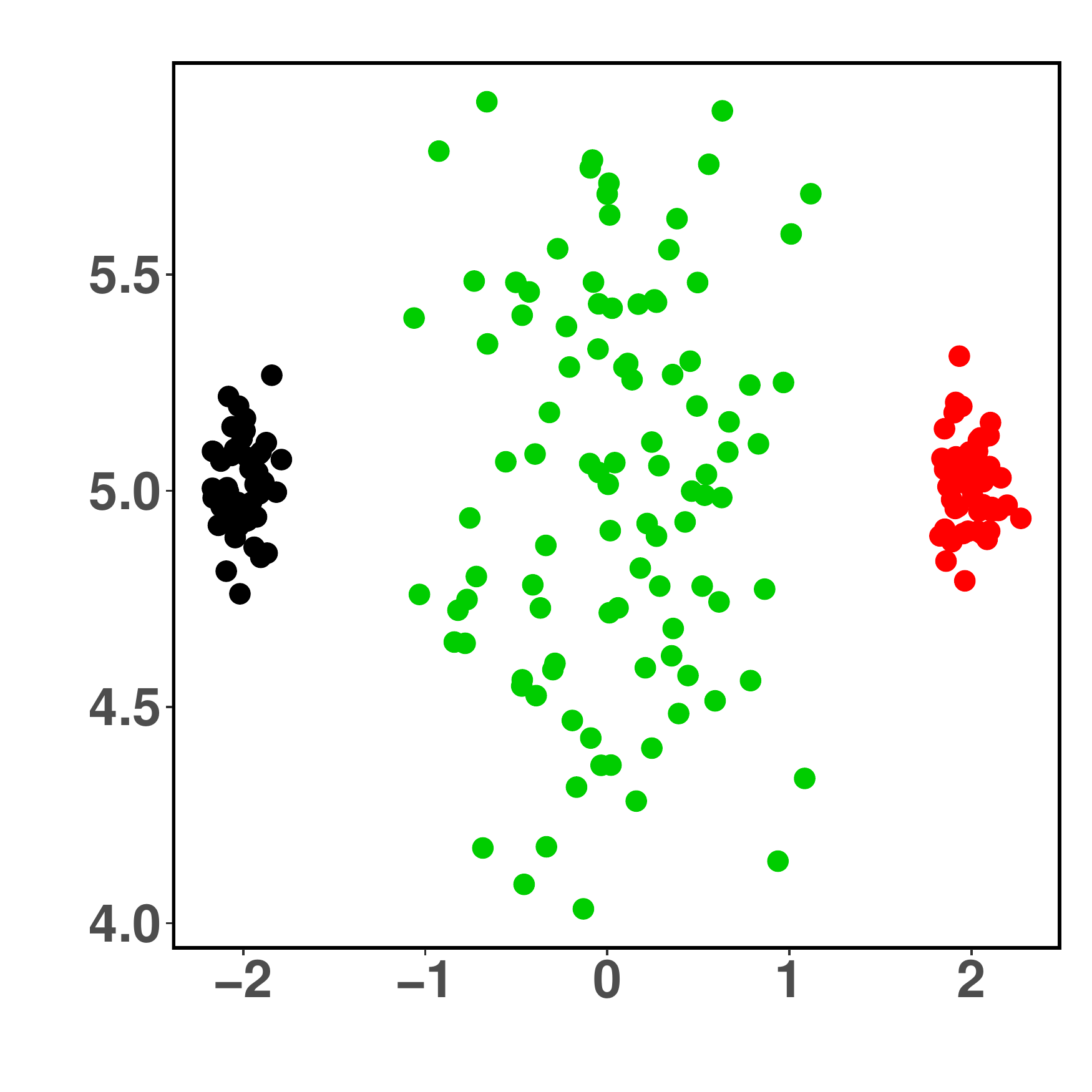}
}
\subfloat[Model 4]{
  \includegraphics[width=35mm,height=35mm]{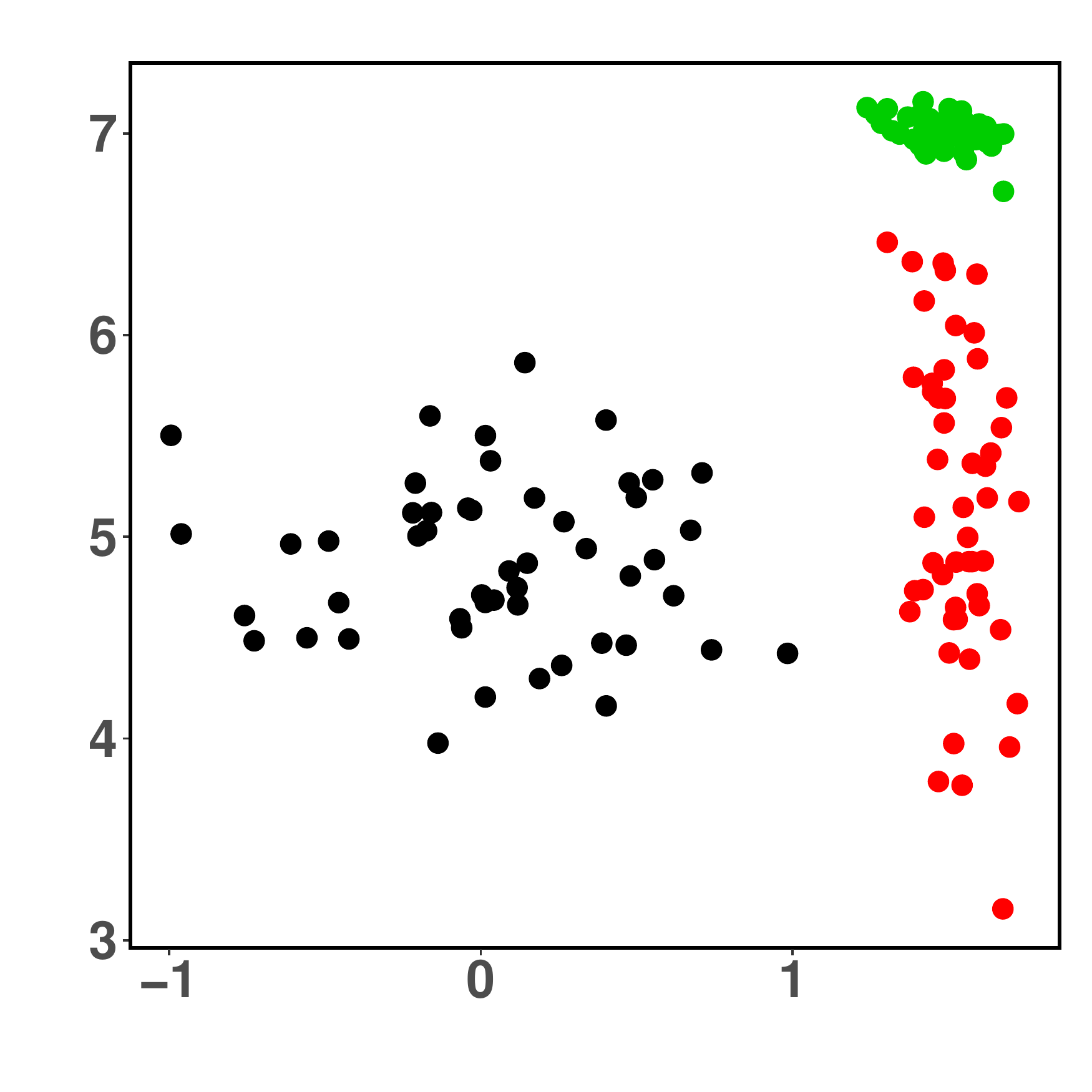}
}
\newline
\subfloat[Model 5]{
  \includegraphics[width=35mm,height=35mm]{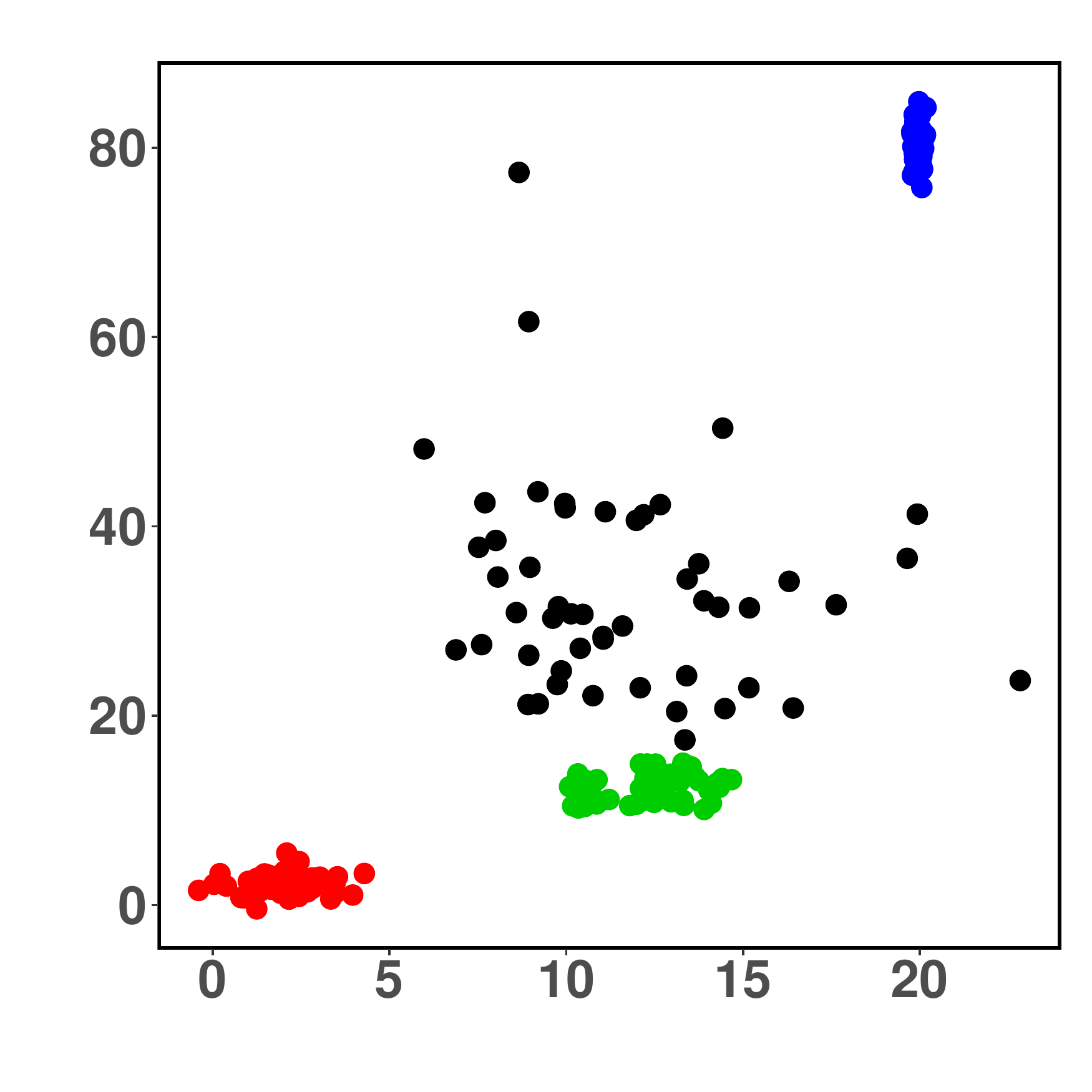}
}
\subfloat[Model 6]{
  \includegraphics[width=35mm,height=35mm]{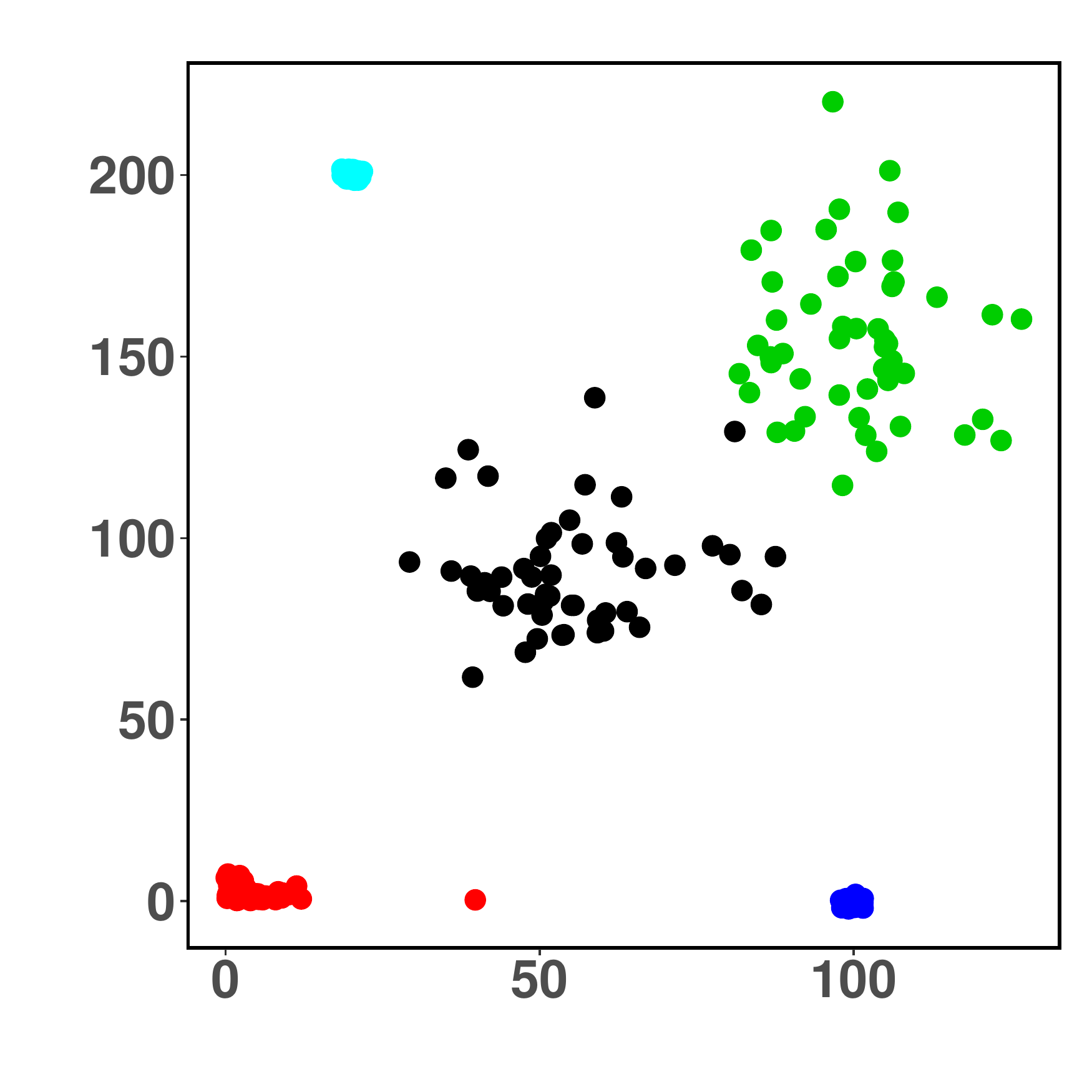}
}
\subfloat[Model 7]{
  \includegraphics[width=35mm,height=35mm]{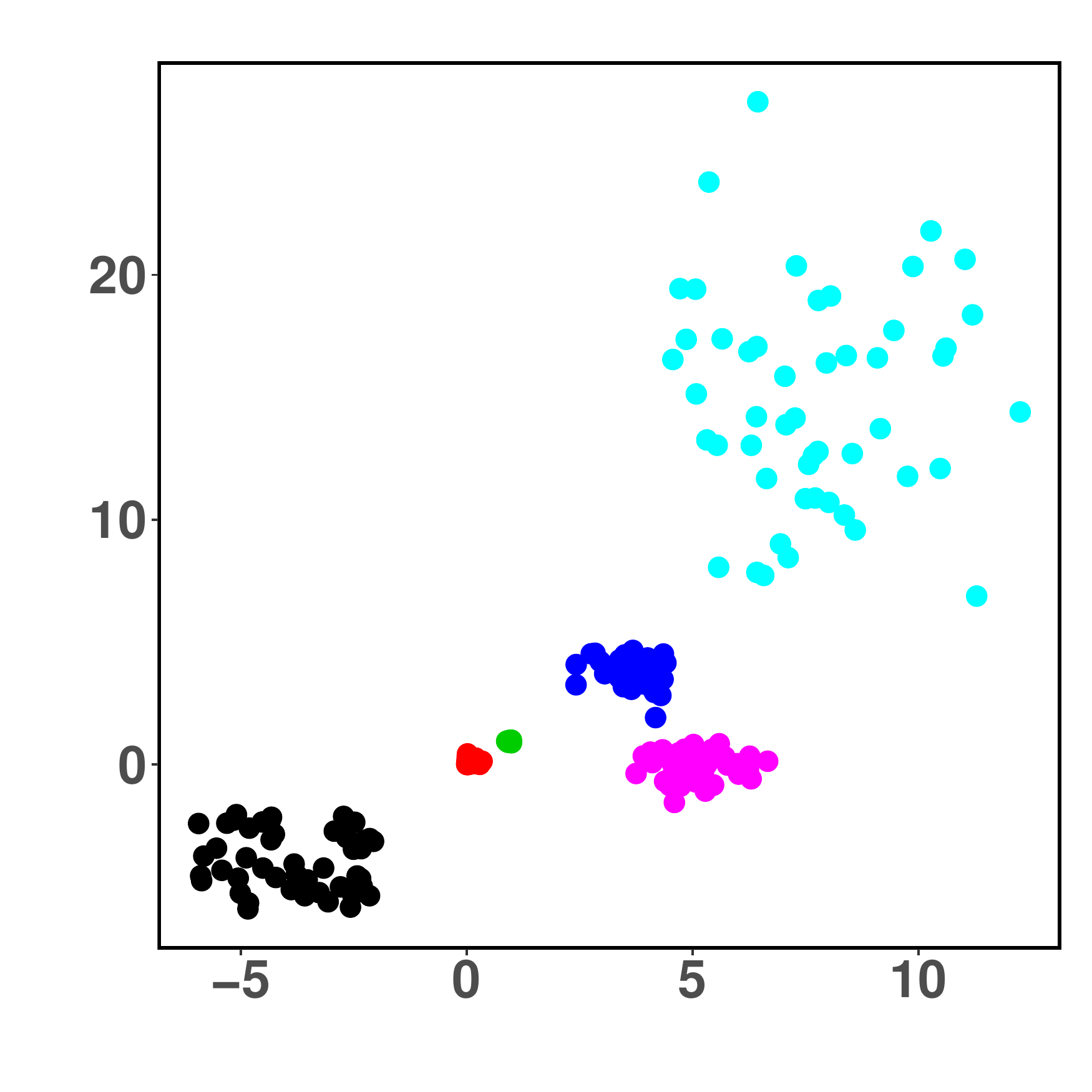}
}
\subfloat[Model 8]{
  \includegraphics[width=35mm,height=35mm]{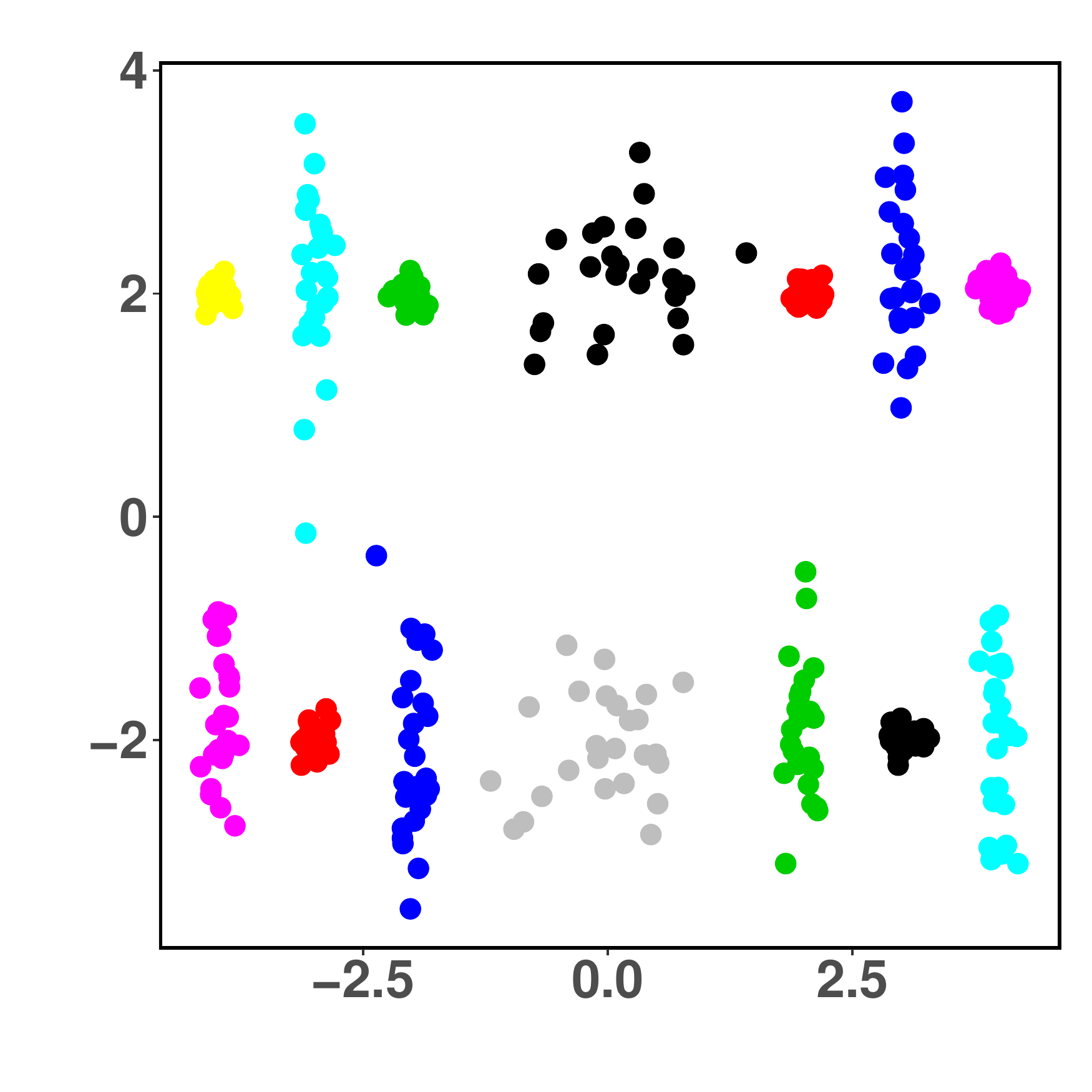}
}
\newline
\rule{-40ex}{.2in}
\subfloat[Model 9]{
  \includegraphics[width=35mm,height=35mm]{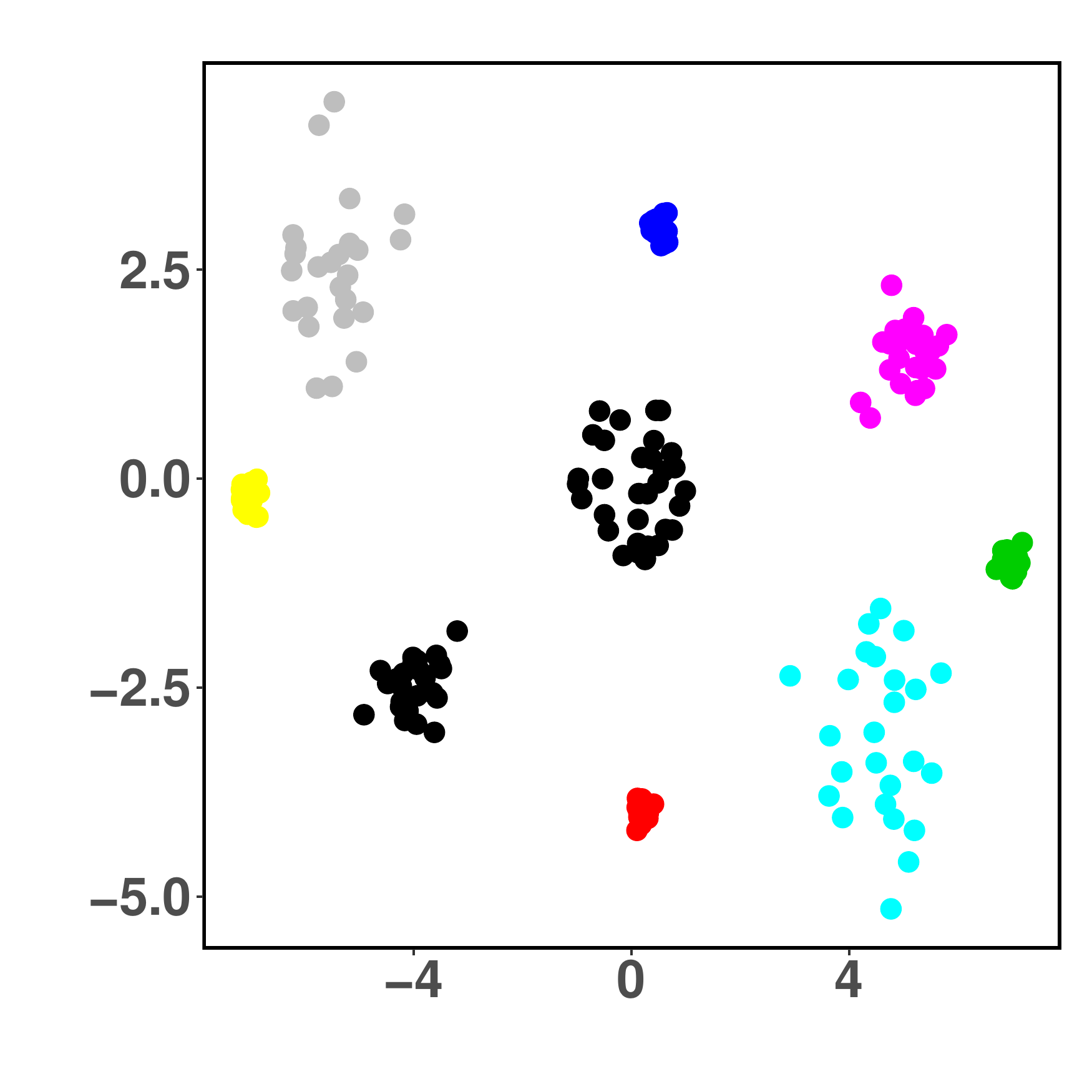}
}
\subfloat[Model 10]{
  \includegraphics[width=35mm,height=35mm]{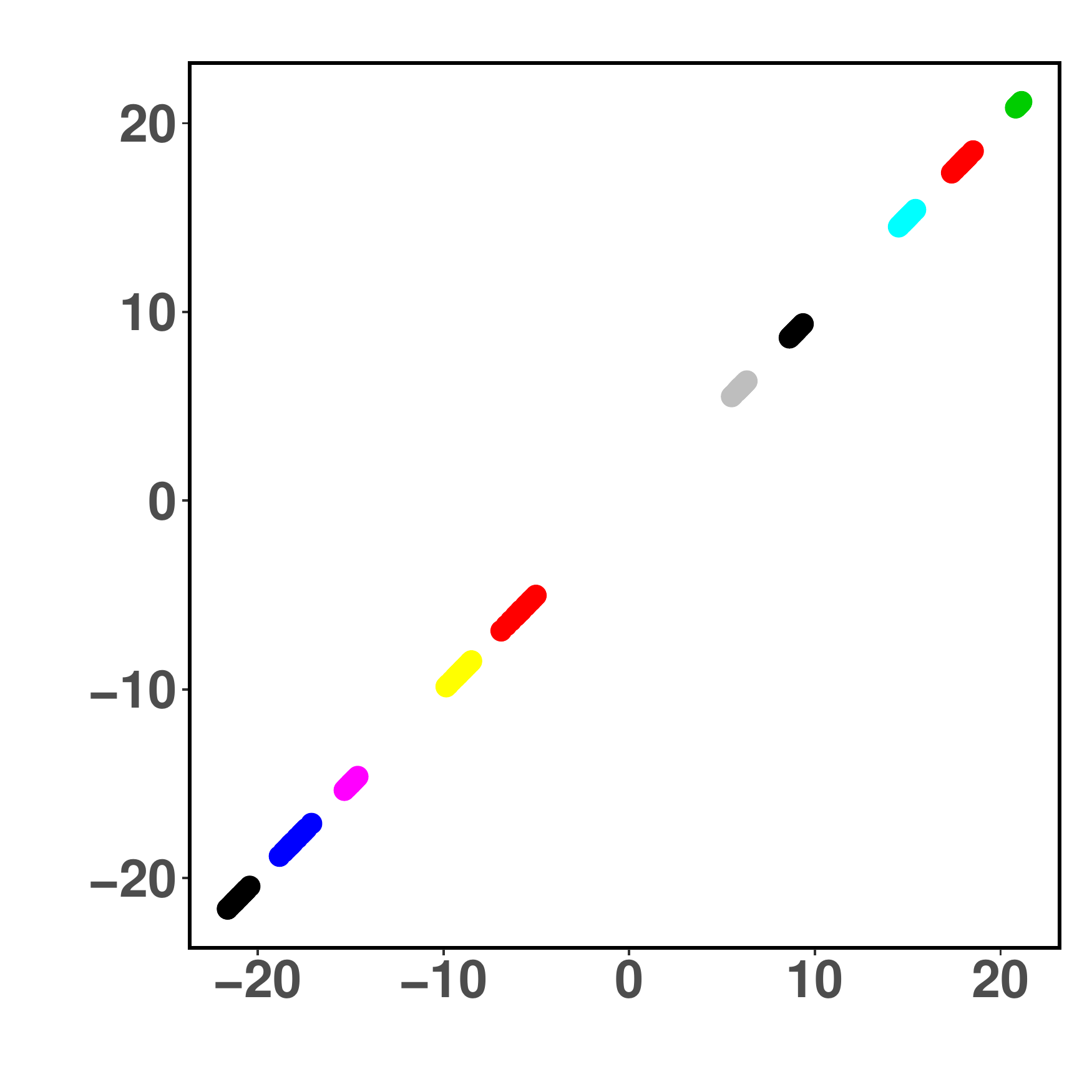}
}
\caption{A dataset generated from each data generating process.}
\label{some}
\end{flushleft}
\end{figure}

$\mathds{R}^p$ is used for the simulation of the data $\Lb$ in experiments, however the proposed algorithms  works with the data from other spaces.   The proposed algorithm also works with general distances, thus, specifying $\Lb$ belongs to some space $S$ characterised by distances $d:S\times S \mapsto \mathds{R}^{+}$, such that $\Lb$, the data is a subset of $S$ is enough for the formalism. Several data generating processes (DGP) having 2 to 14 distinct clusters, different clustering structures, characteristics and complexity have been defined to test the proposed method.  These  DGPs  cover  clustering scenarios having difficulties of various kinds, for instance, clusters from different distributions\textemdash assuming every individual cluster is coming from a single distribution with different variations among observations, equally and unequally sized clusters, clusters from skewed distributions, different types of clusters, for instance, spherical, non-spherical, elongated, close and far away clusters,  i.e., the distance between the means of clusters are varied, nested clusters, clusters with correlated variables, different number of clusters and variables. In this article results from only 10 DGPs covering these features were presented. Figure \ref{some} represents a plot of a dataset each, generated from these DGPs (see Table \ref{defdgp} for a complete list of distributions and parametric choices used to generate data).
\begin{table}[!htb]
 \centering 
\setlength\extrarowheight{-12pt}
\caption{Parameters used in simulations. Simulations include clusters with Gaussian, skew Gaussian, t, non-central t, uniform, $\mathds{F}$, exponential, Beta, and Weibull  distributions.  Dimensions are generated independently and identically. } 
\fontsize{11}{11}\selectfont
 \begin{threeparttable}
\begin{tabular}{ l l l l l l} 
\toprule 
DGP & $k$ & $p$ & Distributions & Cluster size & n\\
\midrule
Model 1  & 2 & 2 & $\textrm{N}((0, 5), I_2)$,  $\mathds{U}(-10, -1)^\ast$  & 100 & 200\\ 
Model 2  & 3 & 2 & $\textrm{N}((0, 5), 0.1I_2)$,  $\textrm{N}((0.5, 5.5), 0.2I_2)$, $\textrm{t}_{25}(5)$, $\textrm{t}_{25}(10)$ & 50,100,50 & 200\\	 
Model 3  & 3 & 2 & $\textrm{N}((-2, 5), 0.1I_2)$, $\textrm{N}((2, 5), 0.1I_2)$, $\textrm{N}((0, 5), 0.5I_2)$   & 50,50,100 & 200\\
Model 4  & 3 & 2 & $\textrm{N}((1.5, 7), 0.1I_2)$,$\textrm{N}((0, 5), 0.5I_2)$, $\textrm{N}((1.5, 5), (0.1, 0, 0, 0.7))$  & 50 & 150\\
	 
Model 5  & 4 & 2 & $\mathds{U}(10, 15)^\ast$, $\mathds{W}(10, 4)^\ast$,$\textrm{t}_{7}(10)$, $\textrm{t}_{7}(30)$ & 50 & 400\\
		 &   &   & $\textrm{N}((2, 2), I_2)$, $\textrm{N}((20, 80), (0.1, 0, 0, 2))$     &    &    \\
		  
Model 6  & 5 & 2 & $\mathds{F}_{(2, 6)}(4)$, $\mathds{F}_{(5, 5)}(4)$, $\chi^2_7(50)$, $\chi^2_{10}(80)$  & 50 & 250\\
 	     &   &   & $\textrm{N}((100, 0), 0.9I_2)$, $\textrm{t}_{40}(100)$, $\textrm{t}_{35}(150)$   &    &    \\
         &   &   & $\textrm{SN}(200, 0.8, 3, 6)$,$\textrm{SN}(20, 0.9, 2, 4)$ &    &    \\        
Model 7  & 6 & 2 & $\mathds{U}(-6, -2)^\ast$, $\textrm{Exp}(10)^\ast$, $\mathds{W}(10, 4)^\ast$, $\textrm{Gam}(15, 2)^\ast$  & 50 & 300\\
 	     &   &   & $\textrm{Beta}(2, 3, 120)^\ast$, $\textrm{SN}(5, 0.6, 4, 5)$,$\textrm{SN}(0, 0.6, 4, 5)$ &    &    \\
Model 8  & 14 & 2 & $\textrm{N}((0, 2), 0.5I_2)$,$\textrm{N}((0, -2), 0.5I_2)$,$\textrm{N}((-4, -2), \Sigma)$  & 25 & 350\\
 	     &   &  & $\textrm{N}((-3, -2), \Sigma)$, $\textrm{N}((-2, -2), \Sigma)$,   $\textrm{N}((2, -2), \Sigma)$  &    &    \\
 	     &   &   &  $\textrm{N}((3, -2), \Sigma)$,  $\textrm{N}((4, -2), \Sigma)$; $\Sigma = (0.1, 0, 0, 0.7)$  &    &    \\
         &   &  & $\textrm{N}((-4, 2), \Sigma_2)$, $\textrm{N}((-3, 2), \Sigma_2)$, $\textrm{N}((-2, 2), \Sigma_2)$   &    &    \\
 	     &   &  &$\textrm{N}((2, 2),\Sigma_2)$, $\textrm{N}((3, 2), \Sigma_2)$, $\textrm{N}((4, 2), \Sigma_2)$: $\Sigma_2= 0.1I_2$)   &    &    \\

Model 9  & 9 & 3 & Unit circle centred at $(0, 0, 0)$ with 33 points & 33, 25 & 233\\
 	     &   &   &  $\textrm{N}((-7, -0.2, -0.2), 0.1I_3)$, $\textrm{N}((0.2, -4, -4), 0.1I_3)$ &    &    \\
 	     &   &   &  $\textrm{N}((0.5, 3, 3), 0.1I_3)$,  $\textrm{N}((7, -1, -1), 0.1I_3)$ &    &    \\
         &   &   &   $\textrm{N}((-5.5, 2.5, 2.5), \Sigma_3)$  &    &    \\
         &   &   &  $\textrm{N}((4.5, -3, -3), \Sigma_3)$; $\Sigma_3 = c(0.6, 0.8, 0.6)$  &    &    \\
          &   &   &  $\textrm{N}((-4, -2.5, -2.5), \Sigma_4)$;  &    &    \\
          &   &   &  $\textrm{N}((5, 1.5, 1.5), \Sigma_4)$; $\Sigma_4 = c(0.4, 0.3, 0.4)$  &    &    \\
Model 10 & 10 & 100 & $\mu_{100} = (-21, -18, -15, -9, -6, 6, 9, 15, 18, 21)^\ast$  & 20, 40, 60, 70,  & 490\\
 	     &   & &  $0.05I_{100}$, $0.1I_{100}$, $0.15I_{100}$, $0.175I_{100}$, $0.2I_{100}$    &    50$\times$ 6&    \\
\bottomrule
\end{tabular}
  \begin{tablenotes}
  \fontsize{8}{8}\selectfont
  \item For the covariance matrices $\Sigma_3$ and  $\Sigma_4$ only main diagonal are mentioned. The off-diagonals entries are zero. The distributions with $^\ast$ shows the other cluster dimensions are distributed same. See Appendix \ref{appone}for full definitions of the DGPs.
  \end{tablenotes}
     \end{threeparttable}
  \label{defdgp}   
\end{table}

\noindent The simulation was done in R language  (\cite{Rcore2019}). Along with the proposed algorithm,  various  widely applicable clustering algorithms have been applied with Euclidean distances, namely $k$-means\textemdash \cite{forgy1965cluster}, partitioning around medoids(PAM)   
 \textemdash  \cite{kaufman1987clustering}, hierarchical clustering algorithms with single linkage\textemdash \cite{sneath1957application}, complete linkage\textemdash \cite{sorensen1948method}, average linkage\textemdash \cite{sokal1958statistical}, Ward method\textemdash \cite{ward1963hierarchical}, McQuitty methods\textemdash \cite{mcquitty1966similarity}, spectral clustering \textemdash \cite{ng2002spectral},  model-based  clustering \textemdash \cite{fraley1998many}), and PAMSIL\textemdash \cite{van2003new} clustering algorithm. For all the hierarchical clustering methods function `hclust()' available with R base package ``stats'' was used. For  Ward's method option ``Wards.D2'' for the method argument of `hclust()' was used. For $k$-means and PAM functions `kmeans()' (with nstart = 100) and `pam()' also available through base  ``stats'', and ``cluster'' (version: 2.0.6, \cite{packagecluster}), were used, respectively. For spectral and model-based clustering the packages "kernlab" (version: 0.9.25, \cite{zeileis2004kernlab}), and ``mclust'' (version: 5.2.3, \cite{fraleymclust}) implementation, available through the  `specc()' and `Mclust()' functions, respectively, were used. For PAMSIL  the standalone C function written by \cite{van2003new} was used. For all the methods, the default settings were used except otherwise stated. $B=50$ datasets were generated for each DGP.  

\par The estimation of number of clusters were also considered from several internal indices with combinations of  several clustering methods. Seven clustering methods namely, $k$-means, PAM and AHC with five linkage methods have been applied with 11 different methods of estimation of number of clusters. In particular,  Calinski and Harabasz (CH) \cite{calinski1974dendrite}, Hartigan   (H) \cite{hartigan1975clustering},  Krzanowski and Lai (KL) \cite{krzanowski1988criterion}, Gap \cite{tibshirani2001estimating},   Jump \cite{sugardocumentation}, Prediction strength (PS)	 \cite{tibshirani2005cluster}, Bootstrap instability (BI) \cite{fang2012selection},  ASW, PAMSIL and HOSil methods to estimate number of clusters were used. In addition Bayesian Information Criterion (BIC)  \cite{schwarz1978estimating} with model-based clustering method as implemented in ``mclust'' to estimate number of clusters was used.

\par The maximum number of clusters allowed for the estimations were 15. 
For H, CH, KL, and Gap  package ``clustersim'' (version: 0.45.2, \cite{walesiak2011clustersim}), for ASW ``cluster'' (\cite{packagecluster}), for PS and BI ``fpc'' (version: 2.1.10, \cite{hennig2010fpc}), and for the Jump method the R code  provided by the author (see \cite{sugardocumentation} were used. Lastly, various transformation powers for the Jump method in simulation were used, particularly, $\hat \delta_k $   using $Y = p/2$, $Y = p/3$, $Y = p/4$, $Y = p/5$, $Y = p/6$ and $Y = p/7$, where $p$ denote the dimensions of the data.

\section{Result and discussion} \label{simres}
\subsection{Clustering performance comparison}
\par The ASW and corresponding ARI values are reported in Table \ref{aswoaswsumma} for the true known clusters. The discussion in this section is based on the average results obtained from simulation. The visual clustering results are presented in the supplementary file. In the following discussion whenever the term ``size of cluster'' is used, I mean to refer the number of observations in clusters. Some general conclusions are evident from the examination of Table \ref{aswoaswsumma}, as given below:

\begin{enumerate}
\item All the methods have produced much higher values of ASW as compared to the true ASW values. 
\item No single method is best for the optimization of the ASW clustering and for obtaining the best ARI values. 
\item It is not necessary that the best ARI values will be achieved for the best ASW clusterings, in fact for the majority of DGPs this is not the case.  
\end{enumerate}

\par Since PAMSIL is the direct competitor of HOSIL for the aim of optimization of the ASW index for clustering, therefore, some specific conclusions related to these two clustering methods drawn from the study are given below:

\begin{enumerate}
\item  Although PAMSIL has produced the best ASW value among the method considered for nearly half of the DGP: 1, 2, 3, 5, 6, and 7. As far as the best ARI is considered PAMSIL only gave the best ARI for two DGPs: 6 and 7.
\item HOSIL has produced the best ASW values for 3 DGPs: 4, 8, and 9.
\item Although PAMSIL has shown better performance in terms of ASW values as compared to HOSIL for about 5 models, it has always produced lesser ARI values in comparison to HOSil, except for models 6 and 7. 
\item The best ARI values obtained were from spectral clustering for model 1, model-based clustering for models 3, 4, and 5, PAMSIL for models 6 and 7, HOSIL for models 2 and 8. However, many methods gave the best ARI values for models 9 and 10. 
\end{enumerate}
 
 \par However, many other interesting conclusions can be drawn about the performance of the clustering methods for each DGP. The  discussion of the results present in the tables for each  DGP individually is included in the supplementary file. 
 
\afterpage{%
\begin{landscape}
\renewcommand{\baselinestretch}{1}
\begin{table}[H]
\fontsize{9}{12}\selectfont
\caption{Clustering quality and the ARI values for all the DGPs against all methods included in simulation for the fixed number of clusters (k).} 
\begin{tabular}{ c c c c c c c c c c c c c c c c c}
\toprule
DGMs & ... & k & true & $k$-means & PAM & single & complete & average & Ward & McQuitty & spectral & model-based & PAMSIL & HOSIL\\
\midrule
\multirow{2}{*}{Model 1} & ASW &  2& 0.6340 & 0.6402 & 0.6400 & 0.5224 & 0.6291 &  0.6323 & 0.6184 & 0.6184 &  0.6259 & 0.6138 & \textbf{0.6461} & 0.6354  \\
						& ARI & &  & 0.8750 & 0.8902 & 0.7761 & 0.9122 &  0.9163 & 0.9040 & 0.9040 &   \textbf{0.9946} & 0.9689 & 0.8845 &  0.9797\\
						\cline{2-15} \\
\multirow{2}{*}{Model 2} & ASW & 3 &  0.5821 & 0.7884 &  0.5845  & 0.7873 & 0.7869 & 0.8028 & 0.7916  & 0.7916 & 0.7441 &  0.7350 &  \textbf{0.8070} & 0.6354 \\
 			   			& ARI & & & 0.4651 & 0.9062 & 0.5103 & 0.4753 & 0.5068 & 0.4839 & 0.4839 & 0.4822 & 0.5409 & 0.4960 & \textbf{ 0.8358}\\
						\cline{2-15} \\
\multirow{2}{*}{Model 3} & ASW & 3 & 0.6750 & 0.6837 & 0.6841 & 0.5342 & 0.5707 & 0.6792  & 0.5359 & 0.5359 & 0.5239  & 0.6758 & \textbf{0.6898} & 0.6799 \\
						& ARI & & & 0.9177 & 0.9250 & 0.6784 & 0.5905 & 0.9554 & 0.6563 & 0.6563 & 0.9476 &  \textbf{0.9984} & 0.9447  &  0.9927 \\
						\cline{2-15} \\
\multirow{2}{*}{Model 4} & ASW & 3&  0.6021 & 0.6357 & 0.6354 & 0.3163 & 0.5217 & 0.6076 & 0.5082 &  0.5082 & 0.5473 & 0.6080 & 0.6390 &  \textbf{0.6716 }\\
						& ARI  & & & 0.8194  & 0.8462 & 0.1377 &  0.5606 & 0.865 & 0.5977 & 0.5977 & 0.8829 &  \textbf{ 0.9834 }& 0.8677 &  0.9415\\
						\cline{2-15} \\
\multirow{2}{*}{Model 5} & ASW & 4 & 0.7388 & 0.7634 &  0.7673    &  0.5941 &  0.6558  & 0.6753 & 0.6630 & 0.6630 &  0.5484 & 0.7279  &  \textbf{0.7650} & 0.7560 \\
						& ARI & & & 0.8830 & 0.9159 &  0.6520&  0.6424 & 0.6425 &  0.6714 & 0.6714 & 0.8100  & \textbf{  0.9787 } & 0.9273   & 0.9659 \\
						\cline{2-15} \\
\multirow{2}{*}{Model 6} & ASW & 5 & 0.8227 & 0.8251 &   0.8250 & 0.6887 & 0.7704 & 0.8205 & 0.8061 & 0.8061 &  0.5877 & 0.7645 & \textbf{0.8259}  & 0.8240 \\
						& ARI & & & 0.9857  & 0.9834 & 0.8159 & 0.9158 & 0.9766 & 0.9573 & 0.9573 & 0.8210 &  0.9405 & \textbf{ 0.9863} & 0.9845 \\
						\cline{2-15} \\
\multirow{2}{*}{Model 7} & ASW & 6 & 0.7479 & 0.7213 &   0.7445   & 0.5654 & 0.5738 &  0.5798 & 0.5730 & 0.5730 &  0.6425   &  0.7330 &  \textbf{0.7485}  &  0.7478 \\
						& ARI & & & 0.7710  &   0.9598  &   0.5967  &  0.2982 &  0.3755 & 0.2938 & 0.2938 &  0.7932 &  0.8176 & \textbf{0.9966 }  & 0.9888 \\
						\cline{2-15} \\
\multirow{2}{*}{Model 8} & ASW & 14 &  0.4331 & 0.7369   &   0.7541    &     0.7508 &  0.5930  &   0.7498    & 0.7187   & 0.7187   & 0.5371  & 0.4354  & 0.6280 &  \textbf{0.7568}  \\
						& ARI & & & 0.9549 & 0.9969  & 0.9958  &  0.7710 &   0.9899  & 0.9295  & 0.9295   &  0.7855  &  0.6396 &  0.7316  &  \textbf{ 0.9985  } \\
						\cline{2-15} \\
\multirow{2}{*}{Model 9} & ASW & 8 & 0.8036 & 0.8036 & 0.8036 & 0.8036  &  0.8014 & 0.8036  &   0.8033  &   0.8033  & 0.6337  & 0.8036  &  0.8037 & \textbf{0.8062}  \\
						& ARI & &  & \textbf{1}  & \textbf{1} & \textbf{ 1 }&  0.9964 &  \textbf{1  }&  0.9993&  0.9993  &   0.8810  & \textbf{ 1} & 0.9983 & \textbf{1}  \\
						\cline{2-15}\\						
\multirow{2}{*}{Model 10} & ASW & 10 &  0.9230 & 0.9101 &  \textbf{0.9230} &  \textbf{0.9230} &\textbf{ 0.9230} & \textbf{0.9230} & \textbf{0.9230 }&\textbf{ 0.9230 }&  0.5846  &  0.8737  &  \textbf{0.9230} & 0.9083  \\
						& ARI & & & 0.9775  &  \textbf{1}  & \textbf{1}  &  0.9998  & 0.9998 &  0.9998 &   0.9998  &  0.8045 &  0.8945 &\textbf{ 1} &  0.9995\\
\bottomrule
\end{tabular}
     \label{aswoaswsumma}
\end{table}
\end{landscape}}

\subsection{Estimation of number of clusters comparison}
\par The results for the estimation of number of clusters ($\hat{k}$) are now presented. The results were analysed using graphical and tabular methods.  First, for each DGP, a complete frequency count table was produced for 1 to 15 clusters, by noticing the number of clusters estimated against each run in the simulation. This produced 10 tables. These tables are provided in the supplementary file to this paper. From these tables a table was produced for only the true number of clusters for each DGP.  This gives  total frequency counts broken down by the DGPs of the correct number of clusters estimated by all the methods  included in the study as reported in  Table \ref{aggregationofresults}. 

{
\renewcommand{\baselinestretch}{1}
\setlength{\tabcolsep}{1.5pt}
\renewcommand{\arraystretch}{1.2}%
\fontsize{9}{9}\selectfont
\begin{longtable}{@{\extracolsep{\fill}}l c c c c c c c c c c c @{}}
 \caption{Frequency table of indication of cluster estimated at correct level for all the indices in combination with all the methods for all DGPs.} \label{aggregationofresults}\\
\toprule
DGPs & M1 & M2 & M3 & M4 & M5 & M6 & M7 & M8 & M9 & M10\\
No.of dims. & 2 & 2 & 2 & 2 & 2 & 2 & 2 & 2 & 3 & 100 \\
No. of clusters &  2 & 3 & 3 &  3 & 4 & 5 & 6 & 14 & 8 & 10 & Overall\\
\midrule
& \multicolumn{10}{c}{H} \\
\cline{2-11}
Single      & 13 & 8 & 13 & 0  &9& 3  & 2  & 0  & 10 &  14 & 72\\
Complete    & 0 & 0 & 0 &  0 & 3 &0 & 6  & 0 &  0 & 0 & 9\\
Average     & 1 & 3 & 10 & 12  & 8& 11 & 0 & 0  & 18 & 0  & 63 \\
Ward        & 0 & 0 & 0 & 1  & 8 & 1 & 2 &  0 & 10 & 0  & 22\\
McQuitty    & 0 & 0 & 0 &  1 & 9 &  1 & 2  & 0  & 10 &  0 &23 \\
kmeans      & 3 & 24 & 1 &  0 & 16  & 16 & 6 & 0  & 2 & 0  &  68 \\
PAM         & 3 & 24 & 1 &  0 & 16  & 17 & 6 & 0  & 2 & 0  &  69\\
&  &  &  &  &  &  &  &  & & & 326\\
\hline
& \multicolumn{10}{c}{CH} \\
\cline{2-11}
Single      & 39 & 8 & 20  &  0 & 0 & 9  & 1 & 24  & 50 & 20  & 171\\
Complete    & 37 & 11 & 1 &  0 & 1  & 0  & 0 &  0 & 36 & 0  &  86\\
Average     & 33 & 2 & 15 & 3  & 0  & 6  & 0 &  23 & 46 & 2  & 130 \\
Ward        & 34 & 4 & 2 &  0 & 0   & 3  & 0 &  6 & 40 &  0 & 89 \\
McQuitty    & 34 & 4 & 2 &  0 & 0   & 3  & 0 &  6 & 40 &  0 & 89\\
kmeans      & 36 & 3 & 0 &  2 & 1   & 11 & 5 &  6 & 2  & 2  & 68   \\
PAM         & 29  & 0 & 0 &  0 & 0  & 0  & 0 &  2 & 24 & 0  &  55\\
&  &  &  &  &  &  &  &  & & & 688\\
\hline
& \multicolumn{10}{c}{KL} \\
\cline{2-11}
Single      & 3 & 4 & 6 & 6  & 5   & 9  & 4 & 0  & 0 & 0  & 37\\
Complete    & 2 & 3 & 4 &  13 & 7  & 11 & 0 & 0  & 2 & 0  &  42\\
Average     & 2 & 5 & 0 &  1 &  9  & 3  & 6 & 0  & 0 & 0  &  26\\
Ward        & 1 & 3 & 8 &  8 &  14 & 1  & 4 & 0  & 0 & 0  & 39\\
McQuitty    & 1 & 3 & 8 &  8 & 14  & 1  & 4 & 0  & 0 & 0  & 39\\
kmeans      & 22 & 5 & 3 &  3 & 10 & 6  & 6 & 0  & 16 & 2  & 73  \\
PAM         & 7  & 6 & 12 & 0  & 8 & 4  & 12 & 0  & 2 &  0 &  51\\
&  &  &  &  &  &  &  &  & & & 307\\
\hline
& \multicolumn{10}{c}{Gap} \\
\cline{2-11}
Single      & 36 & 8 & 20 & 0  & 2   & 9   & 0 &  0 & 50  & 0  & 125 \\
Complete    & 18 & 15 & 17 & 30& 10  & 26  & 0 &  0 & 36 &  0 & 152 \\
Average     & 19 & 7 & 44 & 32  & 10 & 45  & 0 &  0 & 24 &  0 & 181 \\
Ward        & 28 & 14 & 19 & 28  & 9 & 29  & 0 &  0 & 30 & 0  & 157\\
McQuitty    & 26 & 14 & 18 & 28  & 8 & 30  & 0 &  0 & 38 & 0  & 162\\
kmeans      & 11 & 12 & 9 &  18 & 16 & 29  & 0 &  0 & 0 &  0 &  95 \\
PAM         & 12  & 1 & 6 &  13 & 1  & 24  & 2 &  0 & 6 &  0 & 65 \\
&  &  &  &  &  &  &  &  & & & 937\\
\hline
& \multicolumn{10}{c}{Jump} \\
\cline{2-11}
$p$/2 & 1 & 0 & 14 & 0 & 0     & 28 & 19 & 12 & 50  & 26 & 150  \\
$p$/3 & 50 & 0 & 50 & 25 & 10  & 49 & 40 & 12 & 50  & 1 & 287 \\
$p$/4 & 7 & 0 & 50 & 1 & 22    & 50 & 37 & 12 & 50  & 0 & 229  \\
$p$/5 & 0 & 0 & 10 & 0 & 26    & 50 & 0 & 5 & 50 &0  & 141  \\
$p$/6 & 0 & 0 & 0 &  0&  0     & 49 & 0 & 0 & 50  & 0 & 99  \\
$p$/7 & 0 & 0 & 0 & 0 &  1     & 7  & 0 & 0 & 50 & 0 & 58  \\
&  &  &  &  &  &  &  &  & & & 964\\
\hline
& \multicolumn{10}{c}{PS} \\
\cline{2-11}
Single      & 3 & 32 & 2 & 5  & 2    & 6  &  0& 0  & 46 & 46  & 142 \\
Complete    & 2 & 0 & 0 &  0 & 1     & 43 &  0& 0  & 18 & 50  & 114 \\
Average     & 28 & 7 & 0 & 20  & 2   & 43 &  0& 0  & 48 &  46 & 194 \\
Ward        & 44 & 1 & 0 &  29 & 0   & 50 &  0& 0  & 50 & 50  & 224 \\
McQuitty    & 3 & 1 & 0 &  0 & 1     & 40 &  0& 0  & 48 & 48  & 141\\
kmeans      & 50 & 1 & 50 & 42  & 0  & 3  &  0& 0  & 0 &  0 &   146\\
PAM         & 44 & 31 & 1 & 45  & 2  & 50 & 45& 0 & 50 & 50  & 318 \\
&  &  &  &  &  &  &  &  & & & 1279\\
\hline
& \multicolumn{10}{c}{BI} \\
\cline{2-11}
Single      & 0 & 4 & 0 &  6  & 0 & 3   & 3 & 1  & 44 &  0 &61 \\
Complete    & 0 & 0 & 0 &  0  & 0 & 24  & 0 & 1  & 6 &  0 &  31\\
Average     & 2 & 0 & 0 &  3  & 0 & 43  & 0 & 0  & 48 & 0  & 96 \\
Ward        & 28 & 1 & 0 & 26 & 1 & 45  & 0 & 0  & 50 & 0  & 151\\
McQuitty    & 0 & 0 & 0 & 0   & 0 & 31  & 0 & 3  & 30 & 0  & 64\\
kmeans      & 50 & 0 & 50 &23 & 0 & 9   & 0 & 1  & 0 &  0 &  133 \\
PAM         & 32  & 0 & 0 &44 & 7 & 50  & 8 & 1  & 50 & 0  &  192\\
&  &  &  &  &  &  &  &  & & & 728\\
\hline
& \multicolumn{10}{c}{ASW} \\
\cline{2-11}
Single      & 39 & 8 & 20 & 0  & 1 & 10  & 1 & 23  & 50 & 50  & 202\\
Complete    & 50 & 12 & 19 & 22& 1 & 36  & 0 & 2   & 50 & 50  &  242\\
Average     & 50 & 5 & 50 & 36 & 1 & 48  & 0 & 22  & 50 & 50  &  312\\
Ward        & 48 & 5 & 22 & 23 & 2 & 44  & 0 & 11  & 50 &  50 & 255\\
McQuitty    & 48 & 5 & 22 & 23 & 2 & 44  & 0 & 11  & 50 &  50 & 255\\
kmeans      & 50 & 0 & 50 & 48 & 5 & 38  & 17&  7 & 4 &  50 &   269\\
PAM         & 50  & 0 & 50 & 49& 8 & 50  & 40&  9 & 50 &  50 &  356\\
&  &  &  &  &  &  &  &  & & & 1891\\ 
\hline
& \multicolumn{10}{c}{BIC} \\
\cline{2-11}
Model-based & 43 & 1 & 50 & 50 & 33  & 17 & 4 & 0 & 50  & 50 & 298 \\
\hline
PAMSIL & 50 & 0 & 50 & 44 & 9  & 50 & 47 & 1 &  50 & 50 &  351 \\
\hline
HOSil & 50 & 1 & 50 & 44 & 5  & 50 &48  & 21 & 50  & 50 & 369  \\
\bottomrule
\end{longtable}

}

\par  Based on the true number of clusters' determination, the performance of the  H and KL indices was not good for the majority of the models included in the study. The CH  method performed well only for very few models and that was only for one or two clustering methods. The Jump method also estimated correct number of clusters for a very few models (never for Models 2, 8) and very low counts for (Models 5 and 10). The results for the Gap method were below average (less than half times) for the majority of the models included in the analysis. The BI and PS also performed poorly for majority of the models except for a few of them with one or two clustering methods only. Model-based clustering has estimated $\hat{k}$ other than expected for Models 2, 5, 6, 7, and 8.  HOSil clustering estimated  $k$ other than the true $k$ for Models 2 and 5. PAMSIL was also not able to estimate the correct number of clusters for the two models mentioned for HOSil and in addition Model 8. 

\paragraph{Top performing indices} Based on the performance aggregated for all DGPs an overall ranking can be created.   Given  the fact that the total frequency count is 369, HOSil has provided the best performance  for the recovery of  the true number of clusters.  ASW index with PAM clustering method is the second best performing combination with 356 counts. PAMSIL is third in this list with 351 counts. Further, ASW index with average linkage being forth, Model-based clustering with BIC being the fifth, and ASW index with $k$-means clustering is on the sixth rank.

\section{Runtime} \label{runtimeanalysis}
HOSil is implemented in C++ and has an interface for its call in the R language using the Rcpp package (\cite{eddelbuettel2011rcpp}).  The runtime of HOSil is compared with the existing algorithms for all the simulated datasets. The computations were done on the UCL high performance and throughput computing facilities using Legion cluster.  
Table \ref{tab:runtimetable} contains the runtime of HOSil. 
The time reported here does not include the time for distance calculations. The proposed algorithm's performance for retrieving an accurate clustering is good with the advantage that it can also estimate the number of clusters provided the time it takes. In addition, note that the larger $p$ is not a problem for the algorithm (for instance see DGP 10) because it's not represented in distances. Also, note that the time reported in the Table \ref{tab:runtimetable} is to build the full hierarchy, i.e., from 2 to $(n-1)$ clusters. However, the algorithm is implemented in such a way that if $k$ is known, then the algorithm can be stopped at the desired hierarchy level which will significantly reduce the time. 

{
\renewcommand{\baselinestretch}{1}
\setlength{\tabcolsep}{12pt}
\fontsize{9}{11}\selectfont
\begin{table}[!htbp]
\centering
\caption{Time taken by HOSil clustering algorithm for DGPs.}
\begin{tabular}{l l l l l}
\toprule
DGP & $k$ & $p$ &  $n$  & runtime \\
\midrule
Model 1  & 2 & 2 & 200  & 11m  \\ 
Model 2  & 3 & 2 & 200  & 10.6m  \\ 
Model 3 & 3 & 2 & 200  & 10m  \\
Model 4 & 3 & 2 & 250 & 35m \\
Model 5  & 4 & 2 & 200  & 10m  \\ 
Model 6 & 5 & 2 & 250 & 32.5m   \\
Model 7 & 6 &  2 & 300 & 2.7h   \\ 
Model 8 & 14 & 2 & 350 & 6h  \\ 
Model 9  & 9 & 3 & 233  & 43m  \\ 
Model 10 & 10 & 100 & 250 & 1h  \\
\bottomrule
\end{tabular}
\label{tab:runtimetable}
\begin{itemize}[label={}]
\setlength\itemsep{-0.23em}
\item \footnotesize{m = minutes, h = hours. Note that R returns a runtime in seconds. The reported run time here is subject to approximation to minutes and hours. The runtime is the  averaged over all runs. }
\end{itemize}
\end{table}
}

\section{Applications} \label{applications}
\subsection{Tetragonula bee data clustering} \label{tetragonulabees}
In this section, the species delimitation of \cite{franck2004nest} dataset was considered using HOSil. 
The dataset gives the genetic information of 236 species among bees at 13 microsatellite loci from eastern Australia and between Indian and Pacific oceans. The 13 variables are categorical. The purpose of clustering for this data is to find how many bee species are present. 
 \par    The distance measure used was the "shared allele dissimilarity'' particularly designed  for calculation of genetic dissimilarities between species by \cite{bowcock1994high}, and is implemented through the package `prabclus' (\cite{hennig2010prabclus}).  
 
\par
\begin{figure}[H]
\centering
\subfloat[short for lof][]{
 \includegraphics[scale=0.25]{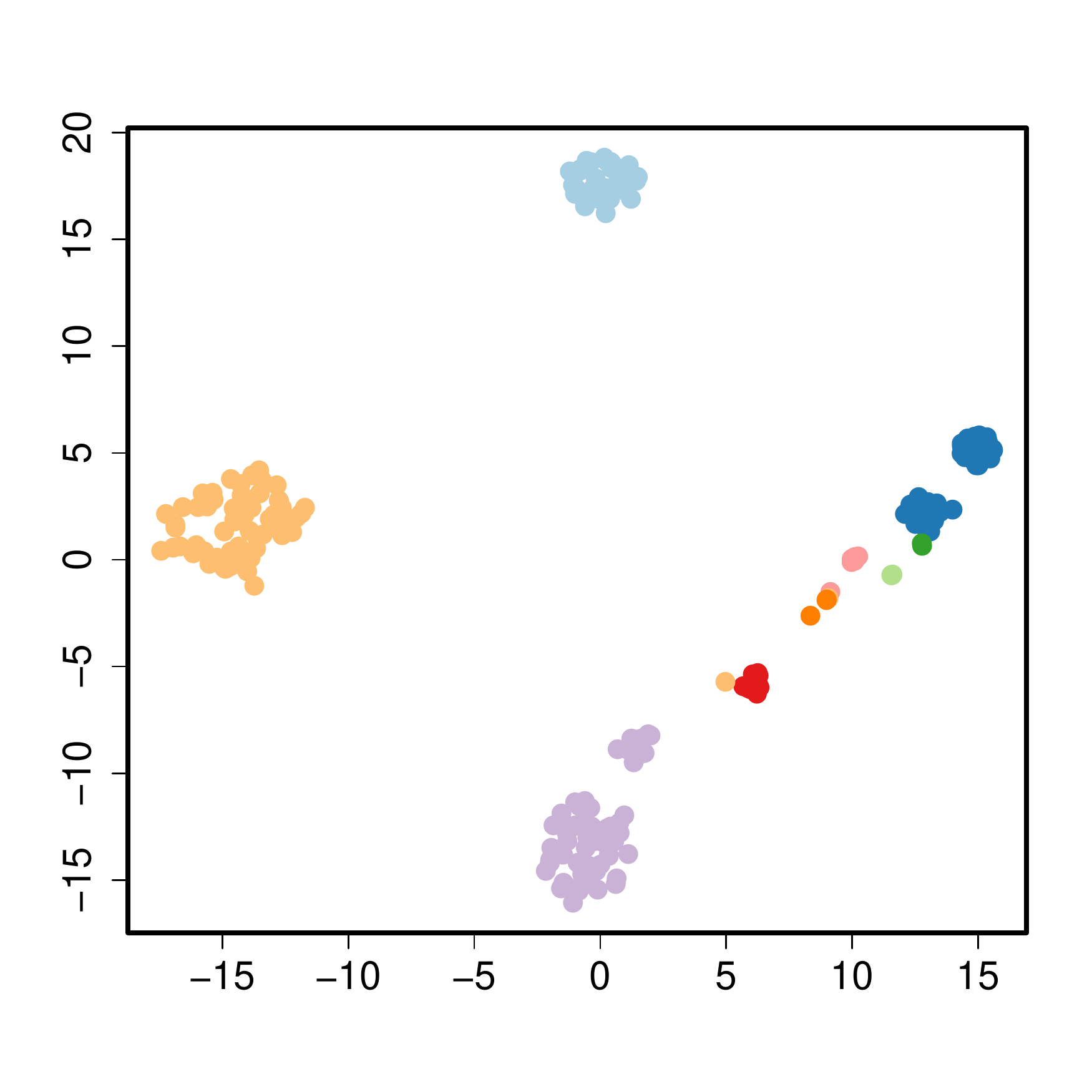} 
}
\subfloat[short for lof][]{
   \includegraphics[scale=0.25]{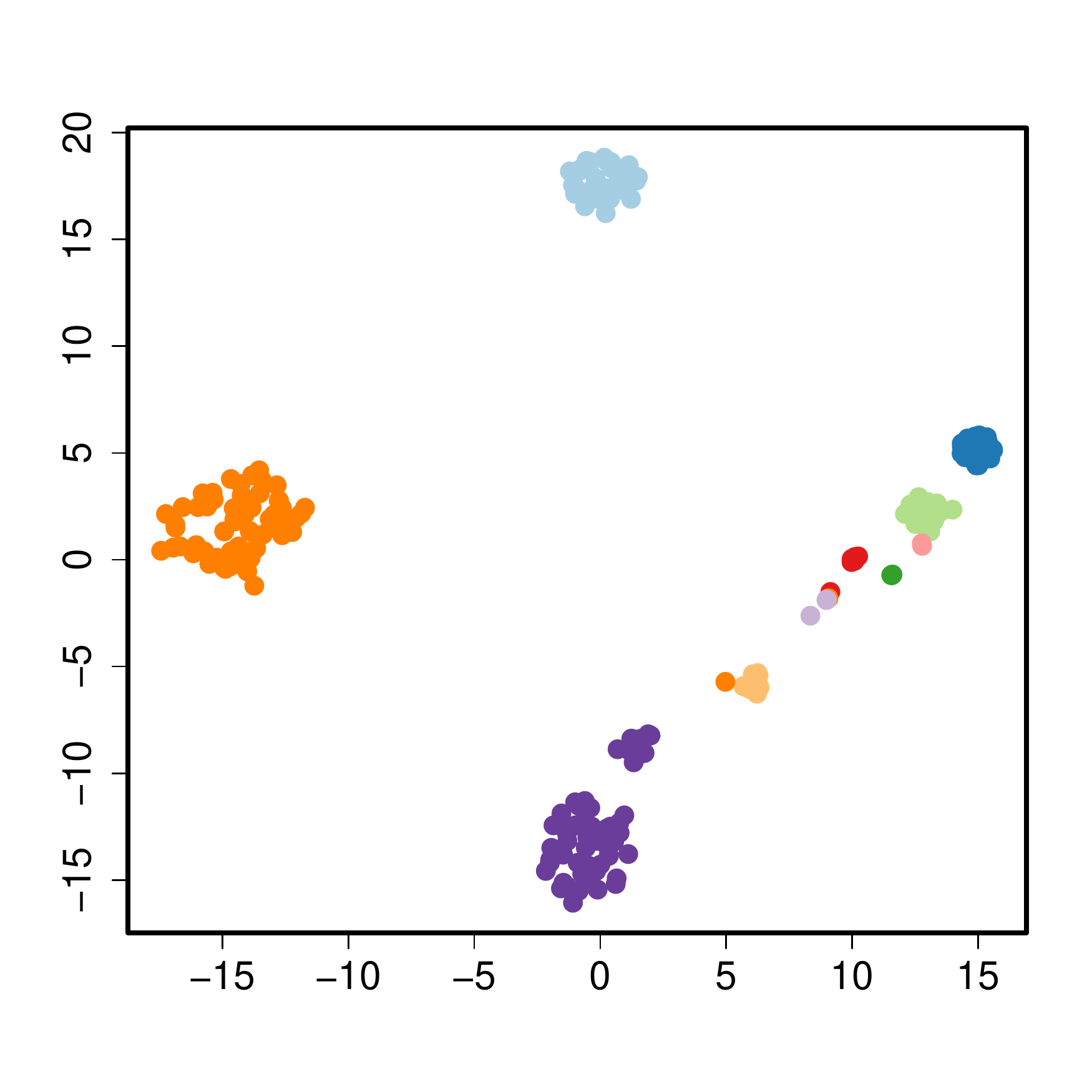} 
}
\subfloat[short for lof][]{
\includegraphics[scale=0.25]{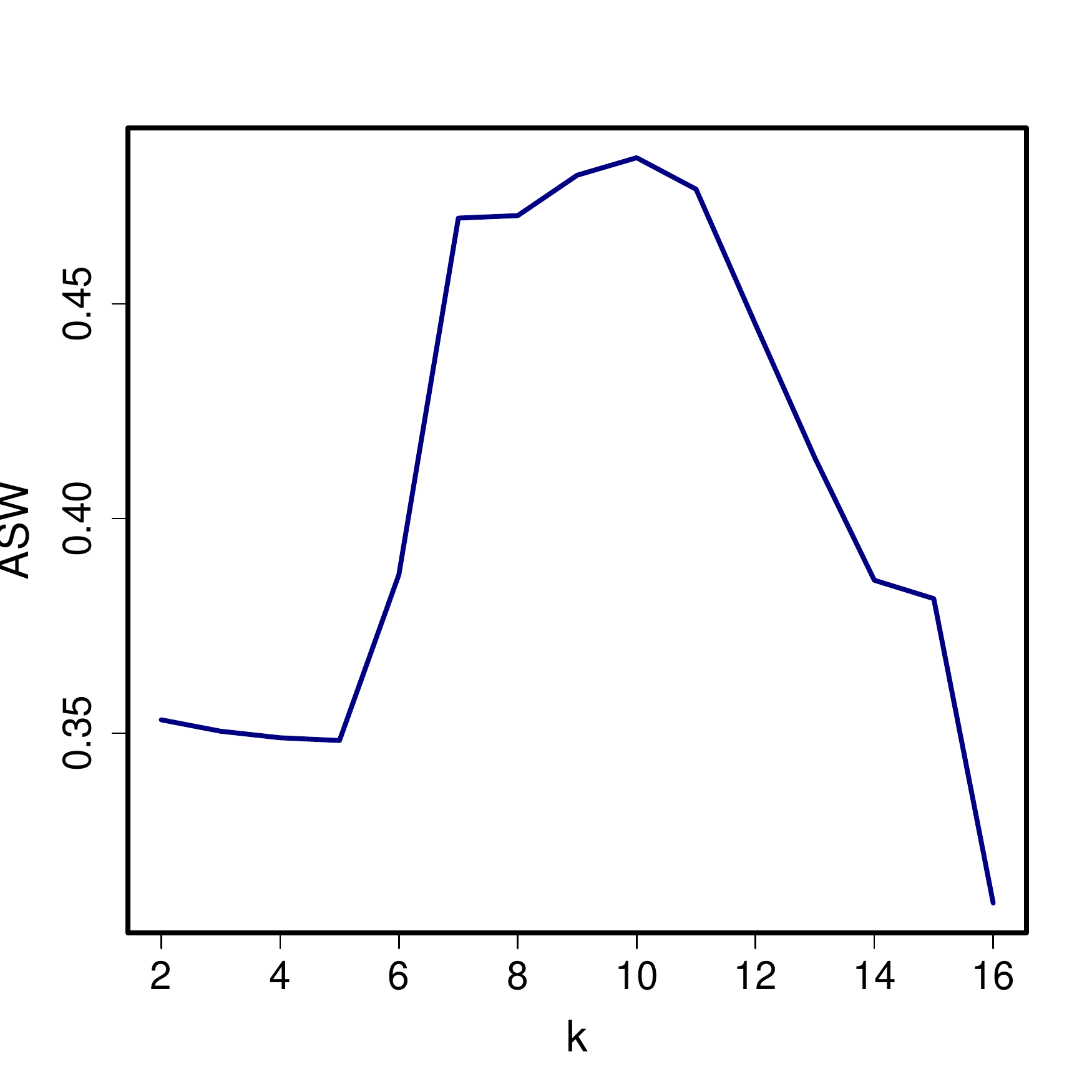} 
  \label{teraK}
}
\caption{t-SNE visualization of  the Tetragonula dataset in two dimensions (a) colours represent the species delimitation provided by  \cite{franck2004nest} (b) colours represent the HOSil clustering results. (c) ASW value obtained from HOSil for 2 to 16 clusters. The maximum value of ASW is obtained at k=10.
}
\label{tetratsne}
\end{figure}

 Since a desirable characteristic to species delimitation could be cluster separation, and the ASW linkage is based on the concept of cluster separation and compactness, therefore, it makes sense to apply HOSil to the bees dataset for clustering.   According to  HOSil the best number of species present in the data is 10. The 4 best ASW values obtained together with adjusted rand index in bold  with the number of clusters were 0.48406(\textbf{0.914795}, k = 10), 0.47999(\textbf{0.91082}, k = 9), 0.47673(\textbf{0.834181}, k = 11), 0.47058(\textbf{0.907194}, k = 8). As evident from the Figure \ref{teraK} as well, the value of the ASW monotonically decreases for the  k before and after k=10(see Figure \ref{teraK} for k = 2 to 16). 
 

\par The  dataset is ploted against the delimitation provided by \cite{franck2004nest}, and HOSil results in two dimensions  with  the t-distributed stochastic neighbor embedding (t-SNE)   available through an R package `Rtsne' (\cite{krijthe2017package})   in Figure \ref{tetratsne}.

 
\subsection{French rainfall data clustering }  \label{franceweatherstation}
Finding spatial or temporal patterns in climate datasets based on statistical techniques is of crucial importance for climatologists. 
The data used in \cite{bernard2013clustering}, available through the software they wrote,   has been considerd for the clustering of french weather into climate regions based on rainfall precipitation maxima observed at the stations. \cite{bernard2013clustering} proposed a clustering algorithm based on a combination of PAM algorithm and a distance measure for geostatistics data called the F-madogram (\cite{cooley2006variograms}).  The data is for 92 French weather stations for the three months of fall, from September to November for 19 years.The purpose of clustering is to find the pattern among stations i.e., spatial clustering based on the fact that if the local conditions at two weather stations are similar, then the two maxima precipitation series at these stations are not independent and the two weather stations should be in one cluster. 
\par The application of the HOSil algorithm to this data offers certain benifts as compared to the algorithm proposed in \cite{bernard2013clustering}.  
All the statistical clustering methods based on averages can not be applied to such datasets due to the fact that the definition of the  distances  for the maxima of time series dependents on the generalised extreme value (GEV) family of distributions (see  \cite{bernard2013clustering}). HOSil does not make use of any kind of cluster representatives like centroids  for clustering. It works with the individual data points. Therefore, it will deal directly with the time series of maxima rather than any kind of averages of these maxima.  In addition, it also offers another advantage that it can estimate number of clusters as well unlike the necessary preprocessing step in \cite{bernard2013clustering}'s method.



\par
\begin{figure}[!htb]
\centering
\subfloat[short for lof][]{
  \includegraphics[clip, trim=0.2cm 0cm 0cm 0cm, width=45mm]{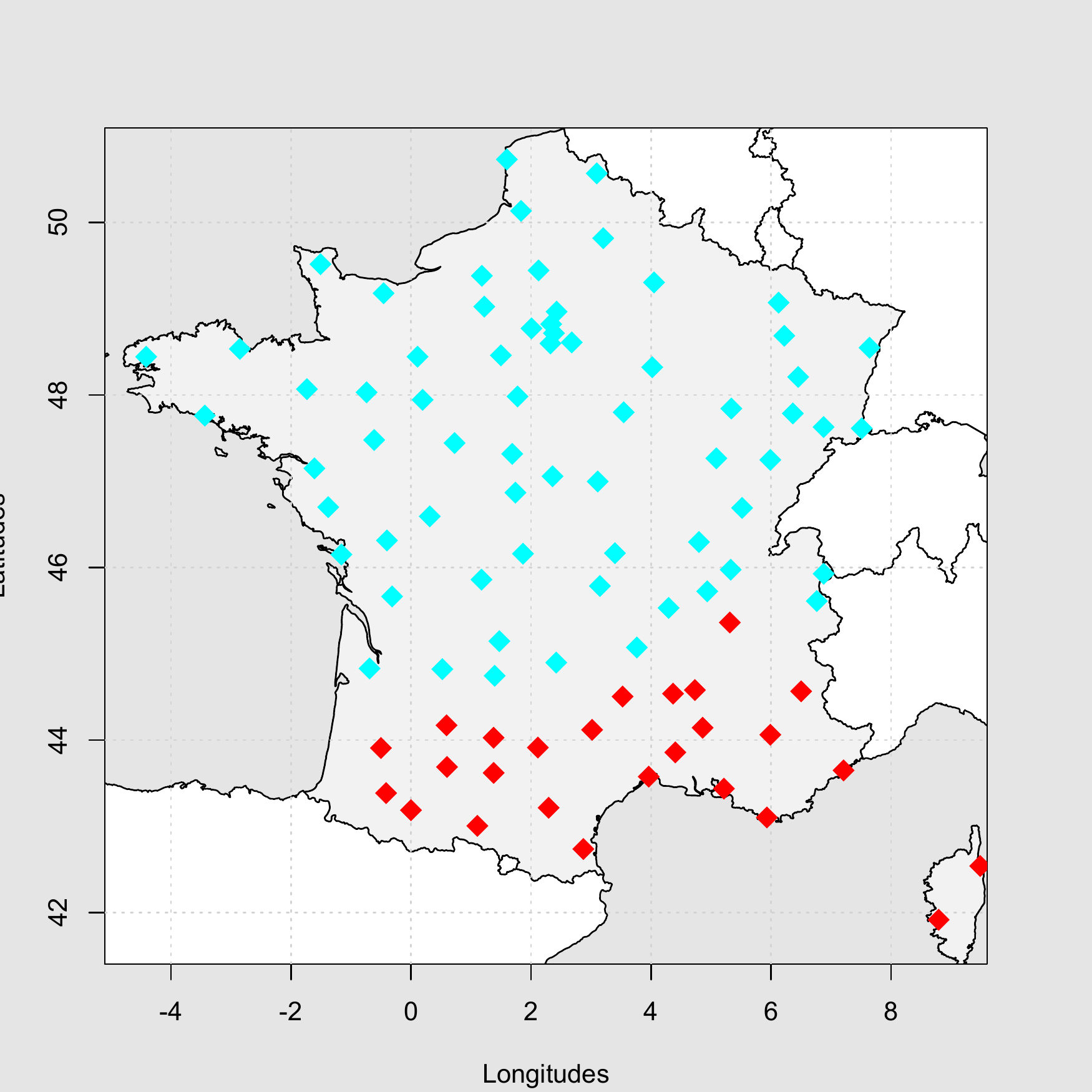}\label{frenchHOSILfiga}
}
\subfloat[short for lof][]{
  \includegraphics[clip, trim=0.2cm 0cm 0cm 0cm, width=45mm]{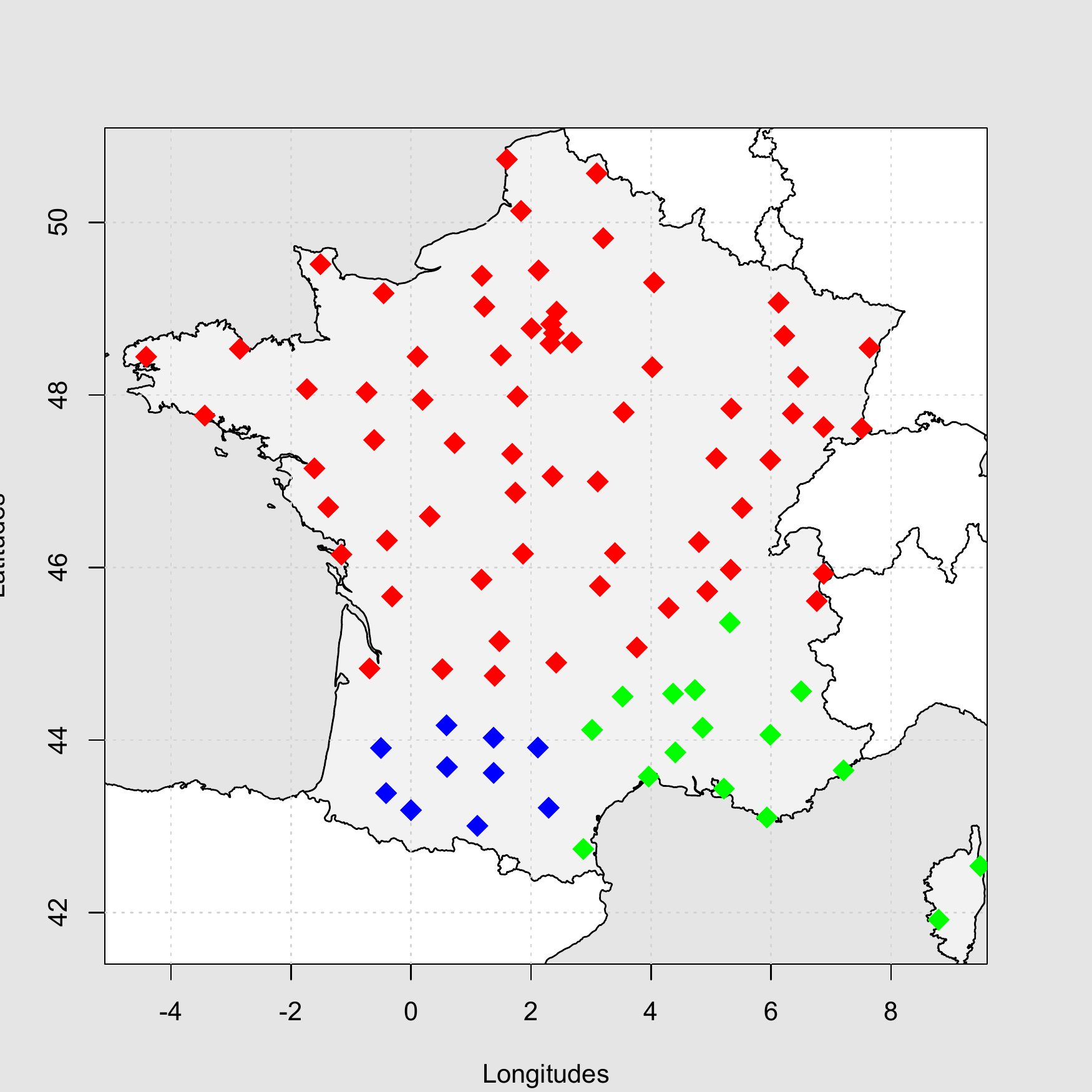}
}
\subfloat[short for lof][]{
  \includegraphics[clip, trim=0.2cm 0cm 0cm 0cm, width=45mm]{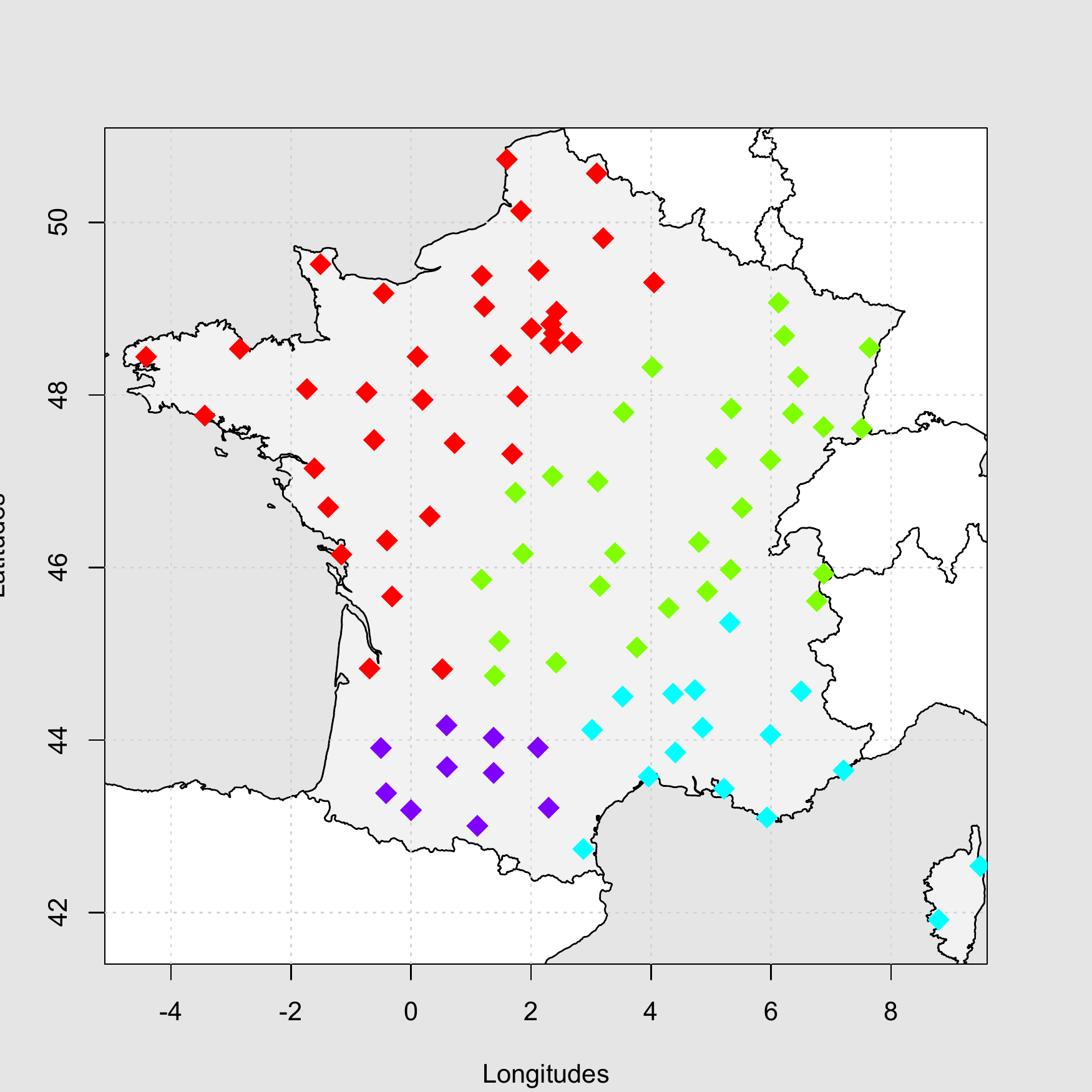}
}
\newline
\subfloat[short for lof][]{
  \includegraphics[clip, trim=0.2cm 0cm 0cm 0cm, width=45mm]{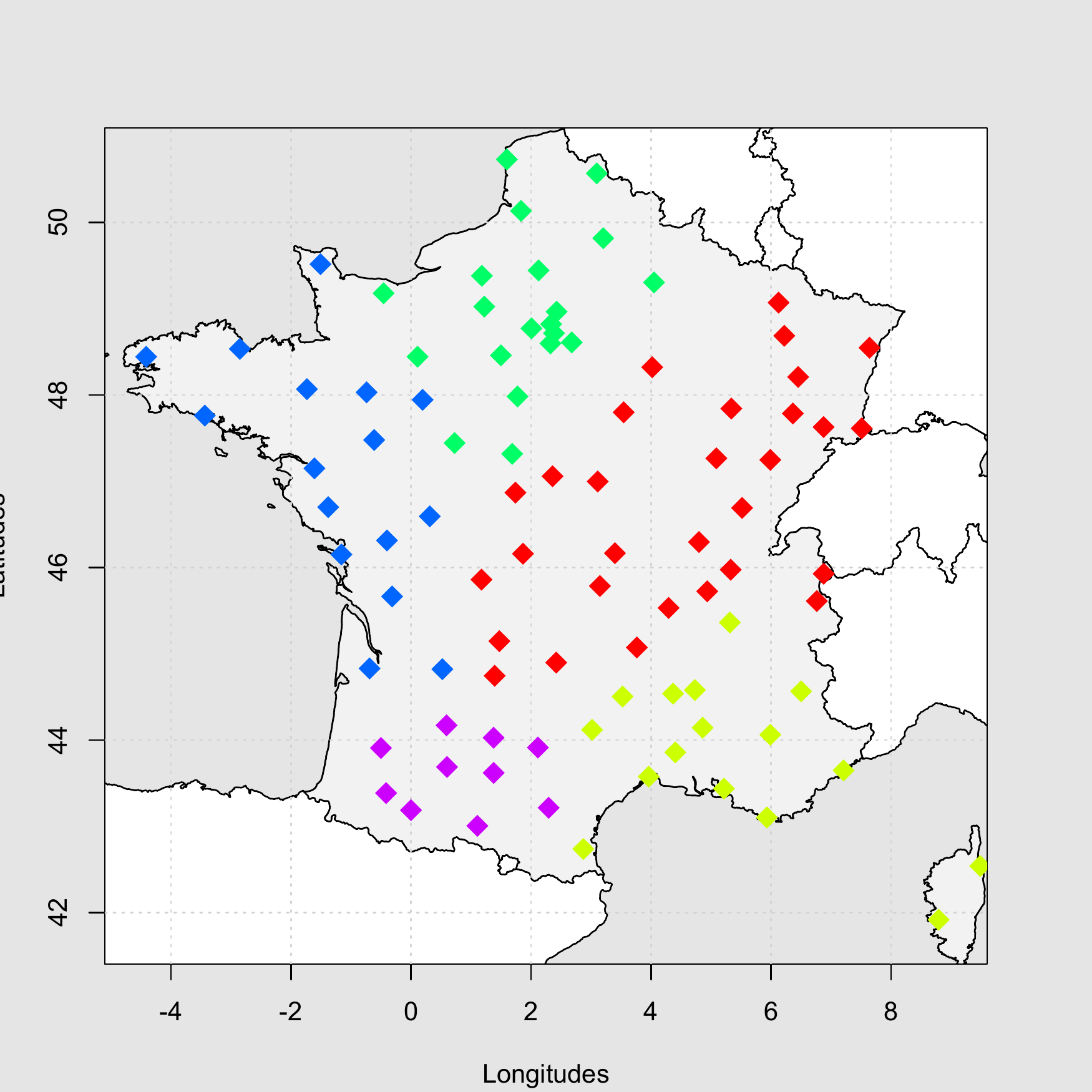}
}
\subfloat[short for lof][]{
  \includegraphics[clip, trim=0.2cm 0cm 0cm 0cm, width=45mm]{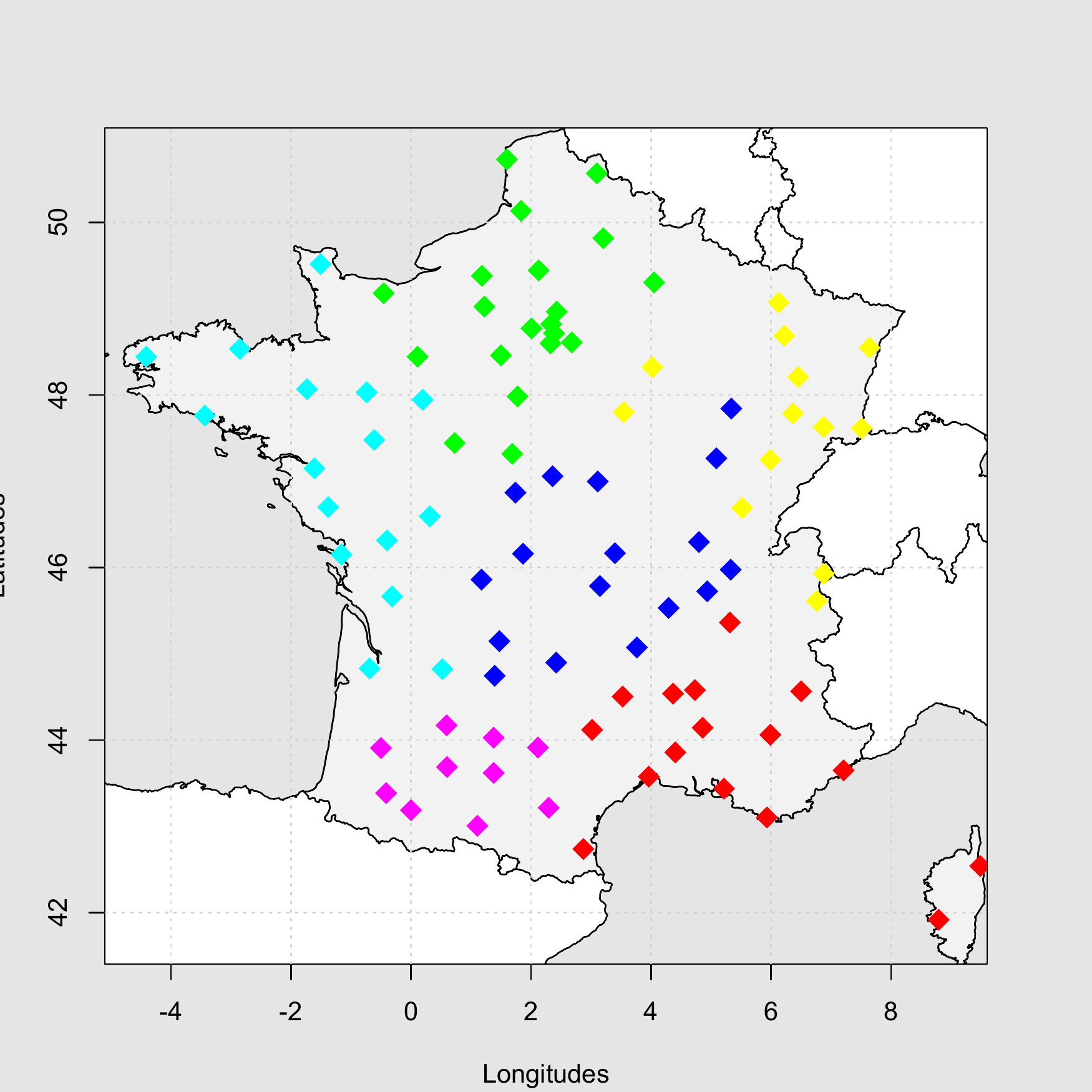}
}
\subfloat[short for lof][]{
  \includegraphics[clip, trim=0.2cm 0cm 0cm 0cm, width=45mm]{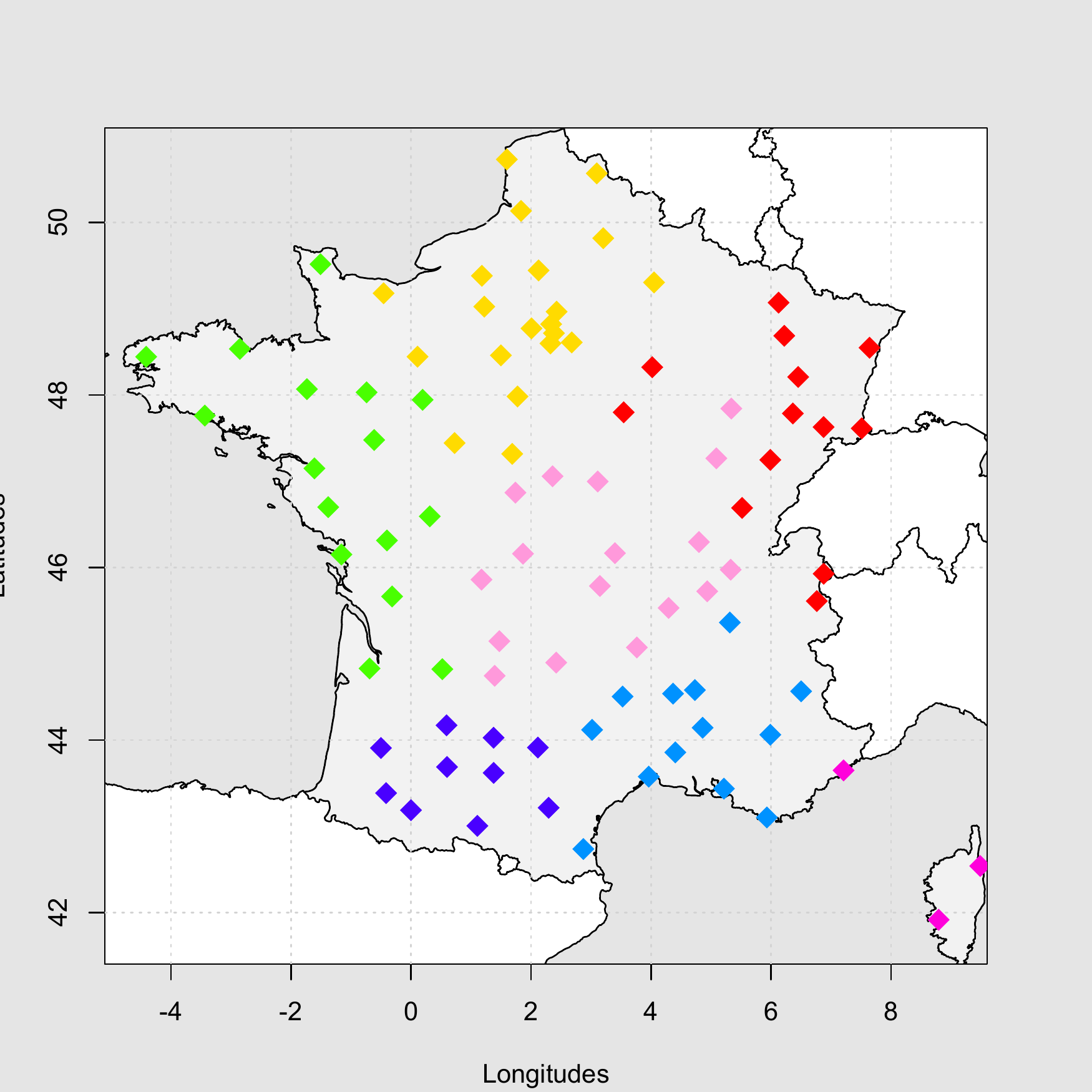}
}
\newline
\caption{Clustering results from HOSil algorithm. Panels from (a) - (f) denote clustering against k=2 to k=7.}
\label{frenchHOSIL}
\end{figure}

 The resulting clusterings from HOSil algorithm are displayed in Figure \ref{frenchHOSIL} for 2 to 7 clusters. For numbers of clusters two HOSil has classified french weather stations in (clock-wise) the east, south and south-west regions of France together in a cluster (red cluster in Figure \ref{frenchHOSILfiga}, say cluster 1), and from the south-west, north, and all the way to the east in the other cluster (blue cluster). The two highest mountain peaks \textemdash the Alps in the east, and the Pyrenees in the south of France are clustered together (cluster 1).  The two weather stations in Corsica,  which has the third highest mountain peaks, are also classified in this cluster.  
Since PAM looks for equally sized clusters, \cite{bernard2013clustering} got an almost equal number of weather stations in the north and south clusters, dividing France along the Loire valley line. HOSil for three clusters further separates cluster 1 in south-east and south-west fashion i.e., isolating the regions with the Alps and Pyrenees. For four clusters,  HOSil has further isolated central France from the northern region. For the five clusters the upper northern region is further divided into  west to north-west, and from north-west to the north regions. 

\par In terms of the number of clusters, the highest ASW value was obtained for k=2 (0.1390), indicating the strongest weather pattern, meaning that the climates of these two regions differ most significantly in the country. The second best ASW value was achieved for  k=3(0.1251). After that, the strongest climate pattern was observed for k=7(0.1166). The fourth best ASW was obtained for k=6(0.1135). 


\subsection{Identification of cell population} 
Identification of cell types in several tissues and organs from the mass of heterogeneous cells is an important task in cell biology.  This is considered as the first step in the biological analysis of the most recently known technique,   single-cell RNA sequencing (scRNA-seq). It is a way to analyse an individual cell and it can inform how each cell is different from the other. The data by \cite{goolam2016heterogeneity}  has been considered for clustering using HOSil. This data is the 
study of pre-implantation development of 124 cells of mouse embryos. There are 5 distinct cell types. 2-cell(16 samples), 4-cell(64 samples), 8-cell(32 samples), 16-cell(6 samples), and 32-cell(6 samples). 


\par
\begin{figure}[H]
\centering
\subfloat{
\includegraphics[width=50mm]{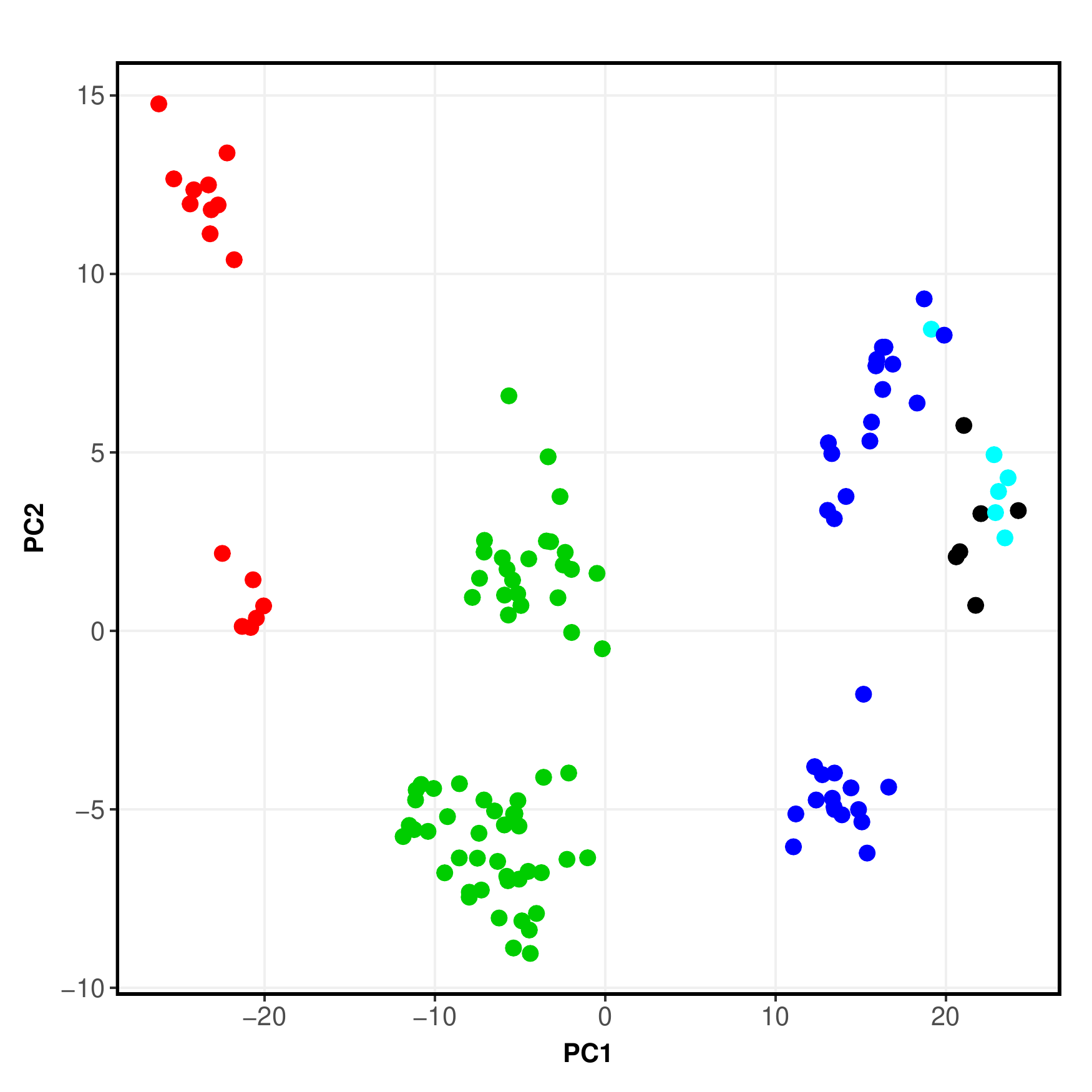}}
\subfloat{
\includegraphics[width=50mm]{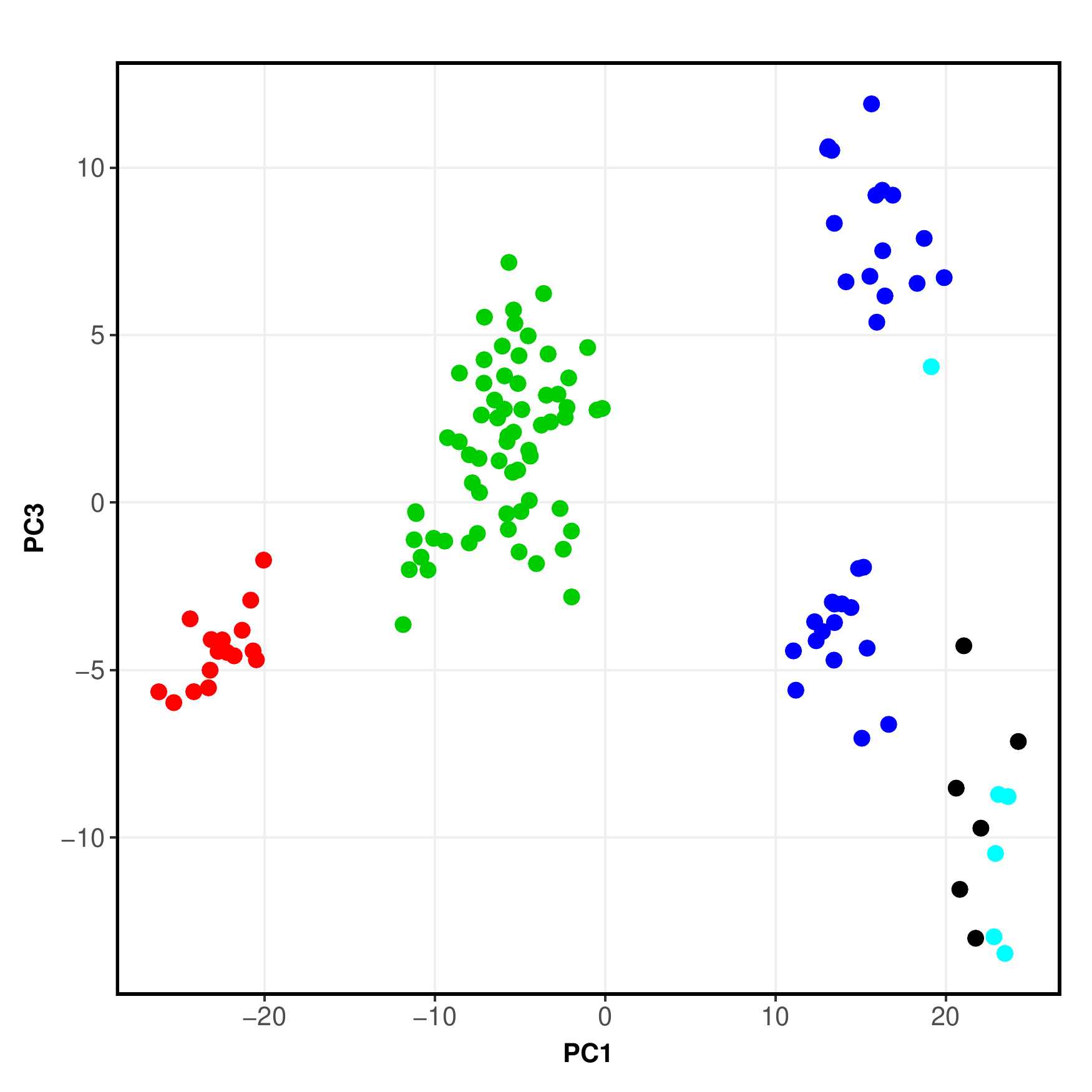}}
\caption{Data plots of first three principal components. Shown in \textcolor{red}{red}    are  2-cell, \textcolor{green}{green}   are 4-cell, \textcolor{blue}{blue}   are 8-cell, \textcolor{black}{black}   are 16-cell and \textcolor[rgb]{0,0.9,0.9}{light blue}  are 32-cell stages. }
\label{goolampca}
\end{figure}

 For the low-level analysis of data, i.e, quality control, normalization and dimensionality reduction,  the R libraries ``scater'' \cite{scater} and ``scran'' \cite{scran} were used which are available through Bioconductor for scRNA-seq data. Euclidean distances between cells to perform clustering were used. The first three principled components of the dataset is plotted in Figure \ref{goolampca}. The colour represents the true cell classification by the authors.  

\par Table  \ref{tab:Goolamtable} shows the results for this data. It is evident from the table that all the methods have produced much higher values of the ASW as compared to the true values. The spectral clustering method gave very low value of ASW. $k$-means, PAM, and model-based clustering have also produced lower values of ASW as compared to other methods with very low ARI values.  These methods were also not able to determine the correct number of clusters. The average linkage, Wards clustering, and HOSil methods have produced the highest ASW values with the best ARI values (see Table \ref{tab:Goolamtable} for values). The average linkage, Wards clustering, and HOSil methods have also estimated correct numbers of clusters. 

{
\renewcommand{\baselinestretch}{1}
\begin{table}[!ht] \centering 
  \caption{\cite{goolam2016heterogeneity} data clustering results.} 
    \fontsize{11}{11}\selectfont
\begin{tabular}{@{\extracolsep{-.1pt}}lcc|ccc} 
\hline 
\hline 
&  \multicolumn{2}{c|}{ True k}   & \multicolumn{3}{c}{ Estimated k}  \\ 
\cline{2-6}
Methods & ASW  & ARI  & ASW  & ARI & $\hat{k}$    \\
\hline
True label & 0.4905 &  \\  
k-means &  0.5502 & 0.5439 & 0.5995 & 0.8831 & 3 \\ 
PAM &  0.5502 & 0.5439 &  0.6365 &  0.8602 & 4\\ 
average & 0.6668 & 0.9097 & 0.6668 & 0.9097 & 5 \\ 
Ward's &  0.6668 & 0.9097 & 0.6668 & 0.9097 & 5 \\ 
model-based & 0.5502 & 0.5439 & 0.6365  & 0.8602 & 4\\ 
spectral & 0.2617 &  0.6365 & 0.3290 & 0.8602 & 4\\ 
BIC-mb &  - & - & 0.5925 & 0.4750 & 8 \\ 
HOSil & 0.6668 &  0.9097 & 0.6668 & 0.9097 & 5\\
\hline
\end{tabular} 
\label{tab:Goolamtable}
\end{table}
}
\section{Final remarks and future work} \label{con}
In this paper, a new AHC methodology was introduced, implemented through the HOSil algorithm. The approach was based on the development of the linkage criterion based on the ASW index. A monte carlo simulation of 11 procedures for determining the number of clusters paired with 7 clustering methods was conducted. For each of these combinations, 10 data structures were used.

\par HOSil can correctly identify a variety of clustering structures other than those presented here. Several other real and artificial structures were used to experiemnt. For instance, HOSil can return the desired clustering structures  for the four shapes and smiley data from the R package ``mlbench'' (version: 2.1.1, \cite{leisch2007mlbench}), 2 diamonds, Hepta, and Tetra data from the fundamental clustering problems suite (FCPS)  of \cite{Ultsch}. It can also estimate the correct number of clusters for these datasets except for the smiley dataset. The smiley dataset contains two Gaussian eyes, one trapezoid nose and a parabolic mouth. For this data HOSil has divided the mouth into many small compact clusters.  The reason for this is that the mouth cluster has bigger within cluster distances as compared to other clusters.  In order to get a good ASW value, the within-cluster dissimilarities should be small as compared to between clusters dissimilarities. In case of the wide spread of observations within clusters i.e., large distances between clusters the small within-cluster dissimilarities requirement of ASW index dominates and it splits this bigger cluster into smaller clusters such that rather than having one cluster with bigger within-cluster distances it prefers to split this cluster into nicely compacted smaller clusters. 

\par For this same reason HOSil can not identify clustering structures like Lsun data from FCPS. For the Aggregation (\cite{gionis2007clustering})  dataset it has identified all the clusters correctly except it has divided the moon-like cluster into two clusters, and has combined the two spherical clusters in one cluster for 7-cluster solution. 

\par HOSil can also identify the correct data structures as well as recover the correct number of clusters for the datasets named ``two elongated clusters in three dimensions'',  ``two close and elongated clusters in three dimensions'', and ``three clusters in a microarray-like setting'' in \cite{tibshirani2005cluster}.

\par Statement like HOSIL ``can only produce equally sized spherical clusters'' is not correct. HOSil can handle, to some extent, the variations among within cluster separations, and small sized clusters in the presence of bigger sized clusters. This also depends upon the difference between the means of clusters. However, unfortunately I am unable to offer more exact statements in this regards and quantify these factors. 
 
\par HOSil can not handle overlapping data structures, and if a HOSil clustering is applied to such dataset it will produce well-separated clusters of equal or unequal sizes. A small study to learn the effect of various distance metrics on HOSil has been conducted. In the study, the Manhattan, Euclidean, and Minkowski distances have been included to observe the differences in the clustering results obtained by the proposed algorithm and other existing methods. Overall, from all the methods the optimization performance gained from Minkowski metric is the highest. The ASW values obtained for all the clustering methods showed same trend across DGPs.  The values obtained from the Minkowski metric were greater than the Euclidean metric, and the values obtained from the Euclidean metric were greater than the Manhattan metric.  A more systematic study\textemdash extended to other data spaces, distance metrics and more DGPs\textemdash is needed, in order to be able to make some in-depth analyses and generalized conclusions to some extent.  
 
\par It is worthwhile to look for some fast approximations to implement ASW based linkage criterion or to improve the  HOSil's performance potential computationally for the datasets of bigger sizes. In this regard, I am working to reduce the computational time complexity of the HOSil algorithm. I am also working to extend the dendogram implementation to HOSIL to visualize the full hierarchy produced by the algorithm.

\section*{Acknowledgement}
The author would like to thank Dr Christian Hennig, University College London for providing the helpful comments on the earlier draft of this manuscript. This work is funded by the Commonwealth scholarships. The high performance computing facilities from University College London were used.  
\appendix
\section{Definition of data generating processes} \label{appone}
 \noindent For the data generating processes (DGPs) several probability distributions have been used. I first define the notations for these distributions. Let $N_p(\mu_p, \Sigma_{p\times p})$ represents the $p$-variate Gaussian distribution with mean $\mu_p$ and covariance matrix $\Sigma_{p\times p}$. Let $SN(\zeta, \omega, \alpha, \tau)$ represent a skew Gaussian univariate distribution with $\zeta, \omega, \alpha, \tau$ as location, scale, shape and hidden mean parameters of the distribution respectively. Let $\mathds{U}(a, b)$ represent the uniform distribution defined over the continuous interval $a$ and $b$. Let $t_v$ represent Student's $t$ distribution with $v$ degrees of freedom. Let $t_r(\nu)$ represent the non-central $t$ distribution with $r$ degrees of freedom and $\nu$ be the non-centrality parameter. Let $\textrm{Gam}(\alpha, \beta)$ represent Gamma distribution where $\alpha$ and $\beta$ are shape and rate parameters, respectively. Let $\textrm{NBeta}(v_1, v_2, \lambda)$ represents the non-central Beta distribution of Type-I with $v_1, v_2$ be two shape parameters and  $\lambda$ being the non-centrality parameter.  Let $\textrm{Exp}(\lambda)$ represent the Exponential distribution with $\lambda$ being the rate parameter. Let $\mathds{F}_{(v_1, v_2)}(\lambda)$ represent the non-central $F$ distribution with $v_1, v_2$  degrees of freedom and $\lambda$ be the non-centrality parameter.  Let $\mathds{W}(\tau, \zeta)$ represent the Weibull distribution with $\tau, \zeta$ as shape and scale parameter, respectively.   Let $I_p$ be the identity matrix of order $p$, where $p$ represents the number of dimensions. The DGPs are defined as below.  
\begin{itemize}[leftmargin=*, label={}]
\item \textbf{Model 1} Two clusters of equal sizes are generated in two dimensions coming from different distributions. 100 observations are generated from the Gaussian distribution with identity covariance matrix centred at (0, 5). 100 observations drawn from $\mathds{U}(-10, 1)$ independently along both dimensions. The result is one compact spherical cluster located at the corner of a uniformly distributed  cluster. 
\item \textbf{Model 2} Three clusters of unequal sizes and variations were generated in two dimensions.  Two clusters were generated from independent bi-variate Gaussian distributions with 50 and 100 observations centred at (0, 5) and  (0.5, 5.5), respectively, with covariance matrix as $0.1I_2$ and $0.2I_2$ respectively.  The third cluster with 50 observations was generated from a non-central $t$ distribution  with $t_{25}(5)$ and $t_{25}(10)$ independently. The clusters are of such nature that the generated non-central $t$ cluster has a wider spread than the two Gaussian clusters  which were kept close to each other as compared to the bigger spread cluster.
\item \textbf{Model 3} Three Gaussian clusters in two dimensions of unequal variations. The clusters contains 50, 50, and 100 observations, with covariance matrices as $0.1I_2$, $0.1I_2$ and $0.5I_2$, while the clusters are centred at (-2, 5), (2, 5), (0, 5), respectively. The result is one bigger,  spherical widely spread cluster located between two small, spherical and compact clusters. 
\item \textbf{Model 4} Three Gaussian clusters in two dimensions of equal sizes, different variations and different shapes. 50 observations were generated from (0, 5) with covariance matrix as $0.5I_2$.   50 observations  were generated from Gaussian distribution with means (1.5, 5) with covariance matrix as $\Sigma =
\begin{bmatrix}
  0.1 &  0 \\
  0 & 0.7
\end{bmatrix}$. 50 observations  were generated from (1.5, 7) with co-variance matrix as $0.1 I_2$. The clusters look like a wider spread spherical cluster located left to the vertical cluster and a smaller compact cluster located just at the top of the  vertical cluster. 
\item \textbf{Model 5} Four clusters of equal sizes each having 50 observations were  generated  in two dimensions. One cluster was generated from independent non-central $t$ distributed variables with parameters $t_7(10)$ and $t_7(30)$. One cluster was generated from $\mathds{U}(10, 15)$ independently along both dimensions. One cluster was generated from bivariate normal distribution parametrized by mean (2, 2), and  covariance $I_2$ generated  independently across both dimensions. The fourth cluster is also from independent bivariate Gaussian distributions parametrized by, mean (20, 80) with covariance matrix $\Sigma =
\begin{bmatrix}
  0.1 &  0 \\
  0 & 2
\end{bmatrix}$.
\item \textbf{Model 6} Five clusters in two dimensions from different distributions.  The clusters are parametrized from Chi-squared, $\mathds{F}$, $t$, Gaussian and skewed Gaussian  distributions as  $\chi^2_7(50)$ and $\chi^2_{10}(80)$, $\mathds{F}_{(2, 6)}(4)$ and  $\mathds{F}_{(5, 5)}(4)$, $t_{40}(100)$ and $t_{35}(150)$, $N((100, 0), 0.9I_2)$, $SN(20, 0.9, 2, 4)$ and $SN(200, 0.8, 3, 6)$. The clusters contains 50 observations each and were generated independently along both dimensions.
\item \textbf{Model 7} 6 clusters from different distributions in two dimensions. The clusters are parametrized as $\mathds{U}(-6, -2)$, $\textrm{Exp}(10)$, $\textrm{Beta}(2, 3, 120)$, $\mathds{W}(10, 4)$, $\textrm{Gam}(15, 2)$ in both dimensions, whereas one cluster from $SN(5, 0.6, 4, 5)$ along first dimension and $SN(0, 0.6, 4, 5)$ across second dimension.  The clusters contains 50 observations each and were generated independently along both dimensions.

\item \textbf{Model 8} Fourteen Gaussian clusters in two dimensions. Two clusters were generated with 25 observations each having  common co-variance matrix as $0.5I_2$ centred at (0, 2) and (0, -2). Six clusters  were generated with 25 observations each having common covariance matrices as $\begin{bmatrix}
  0.1 &  0\\
  0 & 0.7   
\end{bmatrix}$. The clusters are centred at (-4, -2), (-3, -2), (-2, -2), (2, -2), (3, -2), and (4, -2). The remaining six clusters have  common covariance matrix as $0.1I_2$ and 25 observations each centred at  (-4, 2), (-3, 2), (-2, 2), (2, 2), (3, 2), and (4, 2). 
\item \textbf{Model 9} 8 Gaussian clusters surrounding a cluster formed by uniformly distributed points in a unit circle. The unit circle contains 33 observations and is centred at [0, 0, 0]. The 8 Gaussian clusters contain 25 observations each. Four clusters are  centred at (-7, -0.2, -0.2), (0.2, -4, -4), (0.5, 3, 3), and (7, -1, -1) with a common covariance matrix as $0.1I_3$. Two clusters are  centred at (-5.5,  2.5, 2.5) and (4.5, -3, -3) with a common covariance matrix  $\begin{bmatrix}
  0.6 &  0 & 0\\
  0 & 0.8 & 0 \\
  0 & 0 & 0.6  
\end{bmatrix}$.  Two clusters are  centred at (-4,  -2.5, -2.5) and (5, 1.5, 1.5) with a common covariance matrix $\begin{bmatrix}
  0.4 &  0 & 0\\
  0 & 0.3 & 0 \\
  0 & 0 & 0.4  
\end{bmatrix}$.
\item \textbf{Model 10} 10 clusters in 100 dimensions. The clusters are centred at -21, -18, -15, -9, -6, 6, 9, 15, 18, 21. The clusters are in 100 dimensions such that the 100 dimensional mean vectors of these values were generated for all clusters.  The number of observations  for these ten clusters are 20, 40, 60, 70, and 50 each for six of the remaining clusters. The number of observations for the means of clusters were not fix. Any cluster can take  any number of observations from these such that any six clusters have equal number of observations i.e., 50 and the remaining four has different observations each, which is one out of 20, 40, 60, 70 values. The total size of the data is always 490 observations. The covariance matrix for each of these clusters is one out of 0.05$I_{100}$, 0.1$I_{100}$, 0.15$I_{100}$, 0.175$I_{100}$, 0.2$I_{100}$ matrices. The covariance matrix for each cluster was chosen randomly with replacement out of these, such that as a result, all the clusters can have same covariance matrix, two or more clusters can have same covariance matrix  or all of the 10 clusters can have different same covariance matrices. 
\end{itemize}

\bibliographystyle{chicago}
\bibliography{hosil}

\begin{thebibliography}{}

\bibitem[\protect\citeauthoryear{Arbelaitz, Gurrutxaga, Muguerza, P{\'e}rez,
  and Perona}{Arbelaitz et~al.}{2013}]{arbelaitz2013extensive}
Arbelaitz, O., I.~Gurrutxaga, J.~Muguerza, J.~M. P{\'e}rez, and I.~Perona
  (2013).
\newblock An extensive comparative study of cluster validity indices.
\newblock {\em Pattern Recognition\/}~{\em 46\/}(1), 243--256.

\bibitem[\protect\citeauthoryear{Bernard, Naveau, Vrac, and Mestre}{Bernard
  et~al.}{2013}]{bernard2013clustering}
Bernard, E., P.~Naveau, M.~Vrac, and O.~Mestre (2013).
\newblock Clustering of maxima: Spatial dependencies among heavy rainfall in
  france.
\newblock {\em Journal of Climate\/}~{\em 26\/}(20), 7929--7937.

\bibitem[\protect\citeauthoryear{Bowcock, Ruiz-Linares, Tomfohrde, Minch, Kidd,
  and Cavalli-Sforza}{Bowcock et~al.}{1994}]{bowcock1994high}
Bowcock, A.~M., A.~Ruiz-Linares, J.~Tomfohrde, E.~Minch, J.~R. Kidd, and L.~L.
  Cavalli-Sforza (1994).
\newblock High resolution of human evolutionary trees with polymorphic
  microsatellites.
\newblock {\em Nature\/}~{\em 368\/}(6470), 455--457.

\bibitem[\protect\citeauthoryear{Cali{\'n}ski and Harabasz}{Cali{\'n}ski and
  Harabasz}{1974}]{calinski1974dendrite}
Cali{\'n}ski, T. and J.~Harabasz (1974).
\newblock A dendrite method for cluster analysis.
\newblock {\em Communications in Statistics-Theory and Methods\/}~{\em 3\/}(1),
  1--27.

\bibitem[\protect\citeauthoryear{Campello and Hruschka}{Campello and
  Hruschka}{2006}]{campello2006fuzzy}
Campello, R.~J. and E.~R. Hruschka (2006).
\newblock A fuzzy extension of the silhouette width criterion for cluster
  analysis.
\newblock {\em Fuzzy Sets and Systems\/}~{\em 157\/}(21), 2858--2875.

\bibitem[\protect\citeauthoryear{Cooley, Naveau, and Poncet}{Cooley
  et~al.}{2006}]{cooley2006variograms}
Cooley, D., P.~Naveau, and P.~Poncet (2006).
\newblock Variograms for spatial max-stable random fields.
\newblock In {\em Dependence in probability and statistics}, pp.\  373--390.
  Springer.

\bibitem[\protect\citeauthoryear{Eddelbuettel, Fran{\c{c}}ois, Allaire,
  Chambers, Bates, and Ushey}{Eddelbuettel et~al.}{2011}]{eddelbuettel2011rcpp}
Eddelbuettel, D., R.~Fran{\c{c}}ois, J.~Allaire, J.~Chambers, D.~Bates, and
  K.~Ushey (2011).
\newblock Rcpp: Seamless \uppercase{R} and \uppercase{C++} integration.
\newblock {\em Journal of Statistical Software\/}~{\em 40\/}(8), 1--18.

\bibitem[\protect\citeauthoryear{Fang and Wang}{Fang and
  Wang}{2012}]{fang2012selection}
Fang, Y. and J.~Wang (2012).
\newblock Selection of the number of clusters via the bootstrap method.
\newblock {\em Computational Statistics \& Data Analysis\/}~{\em 56\/}(3),
  468--477.

\bibitem[\protect\citeauthoryear{Forgy}{Forgy}{1965}]{forgy1965cluster}
Forgy, E.~W. (1965).
\newblock Cluster analysis of multivariate data: efficiency versus
  interpretability of classifications.
\newblock {\em Biometrics\/}~{\em 21}, 768--769.

\bibitem[\protect\citeauthoryear{Fraley and Raftery}{Fraley and
  Raftery}{1998}]{fraley1998many}
Fraley, C. and A.~E. Raftery (1998).
\newblock How many clusters? which clustering method? answers via model-based
  cluster analysis.
\newblock {\em The Computer Journal\/}~{\em 41\/}(8), 578--588.

\bibitem[\protect\citeauthoryear{Franck, Cameron, Good, Rasplus, and
  Oldroyd}{Franck et~al.}{2004}]{franck2004nest}
Franck, P., E.~Cameron, G.~Good, J.~Y. Rasplus, and B.~Oldroyd (2004).
\newblock Nest architecture and genetic differentiation in a species complex of
  australian stingless bees.
\newblock {\em Molecular Ecology\/}~{\em 13\/}(8), 2317--2331.

\bibitem[\protect\citeauthoryear{Fujita, Takahashi, and Patriota}{Fujita
  et~al.}{2014}]{fujita2014non}
Fujita, A., D.~Y. Takahashi, and A.~G. Patriota (2014).
\newblock A non-parametric method to estimate the number of clusters.
\newblock {\em Computational Statistics \& Data Analysis\/}~{\em 73}, 27--39.

\bibitem[\protect\citeauthoryear{Gionis, Mannila, and Tsaparas}{Gionis
  et~al.}{2007}]{gionis2007clustering}
Gionis, A., H.~Mannila, and P.~Tsaparas (2007).
\newblock Clustering aggregation.
\newblock {\em ACM Transactions on Knowledge Discovery from Data\/}~{\em
  1\/}(1), 4.

\bibitem[\protect\citeauthoryear{Goolam, Scialdone, Graham, Macaulay, Jedrusik,
  Hupalowska, Voet, Marioni, and Zernicka-Goetz}{Goolam
  et~al.}{2016}]{goolam2016heterogeneity}
Goolam, M., A.~Scialdone, S.~J. Graham, I.~C. Macaulay, A.~Jedrusik,
  A.~Hupalowska, T.~Voet, J.~C. Marioni, and M.~Zernicka-Goetz (2016).
\newblock Heterogeneity in oct4 and sox2 targets biases cell fate in 4-cell
  mouse embryos.
\newblock {\em Cell\/}~{\em 165\/}(1), 61--74.

\bibitem[\protect\citeauthoryear{Hartigan}{Hartigan}{1975}]{hartigan1975clustering}
Hartigan, J.~A. (1975).
\newblock {\em Clustering algorithms}.
\newblock New York, USA: John Wiley and Sons.

\bibitem[\protect\citeauthoryear{Hennig}{Hennig}{2015}]{hennig2010fpc}
Hennig, C. (2015).
\newblock {\em fpc: Flexible Procedures for Clustering}.
\newblock R package version 2.1-10.

\bibitem[\protect\citeauthoryear{Hennig and Hausdorf}{Hennig and
  Hausdorf}{2015}]{hennig2010prabclus}
Hennig, C. and B.~Hausdorf (2015).
\newblock {\em prabclus: Functions for Clustering of Presence-Absence,
  Abundance and Multilocus Genetic Data}.
\newblock R package version 2.2-6.

\bibitem[\protect\citeauthoryear{Hruschka and Ebecken}{Hruschka and
  Ebecken}{2003}]{hruschka2003genetic}
Hruschka, E.~R. and N.~F. Ebecken (2003).
\newblock A genetic algorithm for cluster analysis.
\newblock {\em Intelligent Data Analysis\/}~{\em 7\/}(1), 15--25.

\bibitem[\protect\citeauthoryear{Ignaccolo, Ghigo, and Giovenali}{Ignaccolo
  et~al.}{2008}]{ignaccolo2008analysis}
Ignaccolo, R., S.~Ghigo, and E.~Giovenali (2008).
\newblock Analysis of air quality monitoring networks by functional clustering.
\newblock {\em Environmetrics\/}~{\em 19\/}(7), 672--686.

\bibitem[\protect\citeauthoryear{Kaufman and Rousseeuw}{Kaufman and
  Rousseeuw}{1987}]{kaufman1987clustering}
Kaufman, L. and P.~Rousseeuw (1987).
\newblock {\em Clustering by means of medoids}.
\newblock Amsterdam: North-Holland.

\bibitem[\protect\citeauthoryear{Kennedy, Matsuzaki, Dong, Liu, Huang, Liu, Su,
  Cao, Chen, Zhang, et~al.}{Kennedy et~al.}{2003}]{kennedy2003large}
Kennedy, G.~C., H.~Matsuzaki, S.~Dong, W.-m. Liu, J.~Huang, G.~Liu, X.~Su,
  M.~Cao, W.~Chen, J.~Zhang, et~al. (2003).
\newblock Large-scale genotyping of complex dna.
\newblock {\em Nature Biotechnology\/}~{\em 21\/}(10), 1233--1237.

\bibitem[\protect\citeauthoryear{Krijthe, van~der Maaten, and Krijthe}{Krijthe
  et~al.}{201}]{krijthe2017package}
Krijthe, J., L.~van~der Maaten, and M.~J. Krijthe (201).
\newblock Rtsne: T-distributed stochastic neighbor embedding using a barnes-hut
  implementation.

\bibitem[\protect\citeauthoryear{Krzanowski and Lai}{Krzanowski and
  Lai}{1988}]{krzanowski1988criterion}
Krzanowski, W.~J. and Y.~Lai (1988).
\newblock A criterion for determining the number of groups in a data set using
  sum-of-squares clustering.
\newblock {\em Biometrics\/}, 23--34.

\bibitem[\protect\citeauthoryear{Leisch and Dimitriadou}{Leisch and
  Dimitriadou}{2010}]{leisch2007mlbench}
Leisch, F. and E.~Dimitriadou (2010).
\newblock {\em mlbench: Machine Learning Benchmark Problems}.
\newblock R package version 2.1-1.

\bibitem[\protect\citeauthoryear{Liu, Di, Yang, Matsuzaki, Huang, Mei, Ryder,
  Webster, Dong, Liu, et~al.}{Liu et~al.}{2003}]{liu2003algorithms}
Liu, W.-m., X.~Di, G.~Yang, H.~Matsuzaki, J.~Huang, R.~Mei, T.~B. Ryder, T.~A.
  Webster, S.~Dong, G.~Liu, et~al. (2003).
\newblock Algorithms for large-scale genotyping microarrays.
\newblock {\em Bioinformatics\/}~{\em 19\/}(18), 2397--2403.

\bibitem[\protect\citeauthoryear{Lovmar, Ahlford, Jonsson, and
  Syv{\"a}nen}{Lovmar et~al.}{2005}]{lovmar2005silhouette}
Lovmar, L., A.~Ahlford, M.~Jonsson, and A.-C. Syv{\"a}nen (2005).
\newblock Silhouette scores for assessment of snp genotype clusters.
\newblock {\em BMC Genomics\/}~{\em 6\/}(1), 35.

\bibitem[\protect\citeauthoryear{Lun, McCarthy, and Marioni}{Lun
  et~al.}{2016}]{scran}
Lun, A. T.~L., D.~J. McCarthy, and J.~C. Marioni (2016).
\newblock A step-by-step workflow for low-level analysis of single-cell rna-seq
  data with bioconductor.
\newblock {\em F1000Research\/}~{\em 5}, 2122.

\bibitem[\protect\citeauthoryear{Maechler, Rousseeuw, Struyf, Hubert, and
  Hornik}{Maechler et~al.}{2017}]{packagecluster}
Maechler, M., P.~Rousseeuw, A.~Struyf, M.~Hubert, and K.~Hornik (2017).
\newblock {\em cluster: Cluster Analysis Basics and Extensions}.
\newblock R package version 2.0.6.

\bibitem[\protect\citeauthoryear{McCarthy, Campbell, Lun, and Wills}{McCarthy
  et~al.}{2017}]{scater}
McCarthy, D.~J., K.~R. Campbell, A.~T.~L. Lun, and Q.~F. Wills (2017).
\newblock Scater: pre-processing, quality control, normalisation and
  visualisation of single-cell rna-seq data in \uppercase{R}.
\newblock {\em Bioinformatics\/}~{\em 14 Jan}.

\bibitem[\protect\citeauthoryear{McQuitty}{McQuitty}{1966}]{mcquitty1966similarity}
McQuitty, L.~L. (1966).
\newblock Similarity analysis by reciprocal pairs for discrete and continuous
  data.
\newblock {\em Educational and Psychological measurement\/}~{\em 26\/}(4),
  825--831.

\bibitem[\protect\citeauthoryear{Menardi}{Menardi}{2011}]{menardi2011density}
Menardi, G. (2011).
\newblock Density-based silhouette diagnostics for clustering methods.
\newblock {\em Statistics and Computing\/}~{\em 21\/}(3), 295--308.

\bibitem[\protect\citeauthoryear{Ng, Jordan, and Weiss}{Ng
  et~al.}{2002}]{ng2002spectral}
Ng, A.~Y., M.~I. Jordan, and Y.~Weiss (2002).
\newblock On spectral clustering: Analysis and an algorithm.
\newblock In {\em Advances in neural information processing systems}, pp.\
  849--856.

\bibitem[\protect\citeauthoryear{{R Core Team}}{{R Core
  Team}}{2019}]{Rcore2019}
{R Core Team} (2019).
\newblock {\em R: A Language and Environment for Statistical Computing}.
\newblock Vienna, Austria: R Foundation for Statistical Computing.

\bibitem[\protect\citeauthoryear{Recupero}{Recupero}{2007}]{recupero2007new}
Recupero, D.~R. (2007).
\newblock A new unsupervised method for document clustering by using wordnet
  lexical and conceptual relations.
\newblock {\em Information Retrieval\/}~{\em 10\/}(6), 563--579.

\bibitem[\protect\citeauthoryear{Reynolds, Richards, de~la Iglesia, and
  Rayward-Smith}{Reynolds et~al.}{2006}]{reynolds2006clustering}
Reynolds, A.~P., G.~Richards, B.~de~la Iglesia, and V.~J. Rayward-Smith (2006).
\newblock Clustering rules: a comparison of partitioning and hierarchical
  clustering algorithms.
\newblock {\em Journal of Mathematical Modelling and Algorithms\/}~{\em
  5\/}(4), 475--504.

\bibitem[\protect\citeauthoryear{Rousseeuw}{Rousseeuw}{1987}]{rousseeuw1987silhouettes}
Rousseeuw, P.~J. (1987).
\newblock Silhouettes: a graphical aid to the interpretation and validation of
  cluster analysis.
\newblock {\em Journal of Computational and Applied Mathematics\/}~{\em 20},
  53--65.

\bibitem[\protect\citeauthoryear{Schwarz et~al.}{Schwarz
  et~al.}{1978}]{schwarz1978estimating}
Schwarz, G. et~al. (1978).
\newblock Estimating the dimension of a model.
\newblock {\em The Annals of Statistics\/}~{\em 6\/}(2), 461--464.

\bibitem[\protect\citeauthoryear{Scrucca, Fop, Murphy, and Raftery}{Scrucca
  et~al.}{2017}]{fraleymclust}
Scrucca, L., M.~Fop, T.~B. Murphy, and A.~E. Raftery (2017).
\newblock {mclust} 5: clustering, classification and density estimation using
  {G}aussian finite mixture models.
\newblock {\em The {R} Journal\/}~{\em 8\/}(1), 205--233.

\bibitem[\protect\citeauthoryear{Sneath}{Sneath}{1957}]{sneath1957application}
Sneath, P.~H. (1957).
\newblock The application of computers to taxonomy.
\newblock {\em Microbiology\/}~{\em 17\/}(1), 201--226.

\bibitem[\protect\citeauthoryear{Sokal and Michener}{Sokal and
  Michener}{1958}]{sokal1958statistical}
Sokal, R.~R. and C.~D. Michener (1958).
\newblock A statistical method for evaluating systematic relationships.
\newblock {\em Univesity Kansas Science Bulletin\/}~{\em 38\/}(22), 1409--1438.

\bibitem[\protect\citeauthoryear{Sorensen}{Sorensen}{1948}]{sorensen1948method}
Sorensen, T. (1948).
\newblock A method of establishing groups of equal amplitude in plant sociology
  based on similarity of species and its application to analyses of the
  vegetation on danish commons.
\newblock {\em Biologiske Skrifter\/}~{\em 5\/}(4), 1--34.

\bibitem[\protect\citeauthoryear{Sugar and James}{Sugar and
  James}{2003}]{sugardocumentation}
Sugar, C. and G.~M. James (2003).
\newblock Documentation for the \uppercase{R}-code to implement the jump
  methodology in “finding the number of clusters in a data set: An
  information theoretic approach”.
\newblock {\em Marshall School of Business, University of California\/}.

\bibitem[\protect\citeauthoryear{Tibshirani and Walther}{Tibshirani and
  Walther}{2005}]{tibshirani2005cluster}
Tibshirani, R. and G.~Walther (2005).
\newblock Cluster validation by prediction strength.
\newblock {\em Journal of Computational and Graphical Statistics\/}~{\em
  14\/}(3), 511--528.

\bibitem[\protect\citeauthoryear{Tibshirani, Walther, and Hastie}{Tibshirani
  et~al.}{2001}]{tibshirani2001estimating}
Tibshirani, R., G.~Walther, and T.~Hastie (2001).
\newblock Estimating the number of clusters in a data set via the gap
  statistic.
\newblock {\em Journal of the Royal Statistical Society: Series B (Statistical
  Methodology)\/}~{\em 63\/}(2), 411--423.

\bibitem[\protect\citeauthoryear{Ultsch}{Ultsch}{2005}]{Ultsch}
Ultsch, A. (2005).
\newblock Clustering with \uppercase{SOM:U*C}.
\newblock In {\em In Proceeding of Workshop on Self-Organizing Maps, Paris,
  France}, pp.\  75--82.

\bibitem[\protect\citeauthoryear{Van~der Laan, Pollard, and Bryan}{Van~der Laan
  et~al.}{2003}]{van2003new}
Van~der Laan, M., K.~Pollard, and J.~Bryan (2003).
\newblock A new partitioning around medoids algorithm.
\newblock {\em Journal of Statistical Computation and Simulation\/}~{\em
  73\/}(8), 575--584.

\bibitem[\protect\citeauthoryear{Walesiak and Dudek}{Walesiak and
  Dudek}{2017}]{walesiak2011clustersim}
Walesiak, M. and A.~Dudek (2017).
\newblock {\em clusterSim: Searching for Optimal Clustering Procedure for a
  Data Set}.
\newblock R package version 0.47-1.

\bibitem[\protect\citeauthoryear{Ward~Jr}{Ward~Jr}{1963}]{ward1963hierarchical}
Ward~Jr, J.~H. (1963).
\newblock Hierarchical grouping to optimize an objective function.
\newblock {\em Journal of the American Statistical Association\/}~{\em
  58\/}(301), 236--244.

\bibitem[\protect\citeauthoryear{Zeileis, Hornik, Smola, and
  Karatzoglou}{Zeileis et~al.}{2004}]{zeileis2004kernlab}
Zeileis, A., K.~Hornik, A.~Smola, and A.~Karatzoglou (2004).
\newblock kernlab \textemdash an s4 package for kernel methods in
  \uppercase{R}.
\newblock {\em Journal of Statistical Software\/}~{\em 11\/}(9), 1--20.

\end{thebibliography}
\end{document}